\documentclass[aps,prl,twocolumn,superscriptaddress,citeautoscript]{revtex4}
\usepackage{amsfonts}
\usepackage{amssymb}
\usepackage{amsmath}
\usepackage{epsfig}
\usepackage{graphicx}
\usepackage{dcolumn}
\usepackage{bm}
\usepackage{flafter}
\usepackage{float}
\usepackage[normalem]{ulem}
\usepackage{color}
\usepackage{makecell}

\usepackage[final]{pdfpages}

\graphicspath{{figures/}}

\usepackage{siunitx} 
\usepackage{physics} 

\definecolor{jb_color}{cmyk}{1, 0.5, 0, 0}
\definecolor{as_color}{cmyk}{0.7, 0, 1, 0.2}
\definecolor{hs_color}{cmyk}{0.5, 1, 0, 0}
\definecolor{cwh_color}{cmyk}{0, 1, 1, 0}
\definecolor{sectioncolor}{cmyk}{1, 0, 1, 0}

\newcommand{\figSetupCaption}{(a) Schematic illustration of the electronic structure above and below the nematic transition temperature $T_\text{s}$. Fermi surfaces are colored by their dominant
orbital content. For $T<T_\text{s}$, $k_x$ is oriented along the crystalline $a$ axis, where $a>b$. (b) Piezoelectric uniaxial stress apparatus with a platform. (c)Photograph of Sample B, with
contacts attached for measuring resistivity along the sample's length. The lattice directions are indicated. Sample A was prepared similarly, though with its crystal axes rotated by $45^\circ$. (d)
Scanning electron (SEM) micrograph of a cut, made with a focused ion beam, through Sample B and the epoxy layer beneath it. (e) Photograph and (f) SEM micrograph of Sample C, which was prepared in a
Montgomery configuration.}

\newcommand{\figSampleACaption}{The effect of transverse strain. (a) $\rho_{110}(T)$, the resistivity along the $[110]$ direction, for Sample A at various applied strains $\varepsilon_{110}$. The
principal axes of the nematicity in FeSe are the $\langle 100 \rangle$ axes, so this strain is a transverse field to the nematicity. The inset is a schematic of the strain axis.  (b) $T_\text{s}$
versus $\varepsilon_{110}$ for this sample. $T_\text{s}$ is identified as the maximum in $d^2\rho_{110}/dT^2$.  The shaded region is a measure of the width of the transition; it is where
$d^2\rho/dT^2$ exceeds half its maximum value. The line is a fit.}

\newcommand{\figPhaseDiagramsCaption}{Schematic phase diagrams. (a) Schematic stress-temperature phase diagram for the nematicity of FeSe, for stress applied with B\textsubscript{1g} principal axes;
the first-order transition is where the direction of the nematicity flips. (b) The corresponding strain-temperature phase diagram. In the indicated region, the lattice is unstable and breaks up into
twins where, locally, $\varepsilon_{\text{B1g}} = \pm \varepsilon_\text{s}(T)$. }

\newcommand{\figNearTsCaption}{Elastoresistivity near $T_\text{s}$. (a) $\rho_{100}(\varepsilon_{\text{B1g}})$, where $\rho_{100}$ is the resistivity along the $[100]$ direction and
$\varepsilon_{\text{B1g}} \equiv (\varepsilon_{100} - \varepsilon_{010})/2$, of Sample B at various temperatures near $T_\text{s}$. (b) $d\rho_{100}/d\varepsilon_{\text{B1g}}$ for the curves from
panel (a). For $T < T_\text{s}$, $d\rho_{100}/d\varepsilon_\text{B1g}$ becomes nearly constant over the range where the sample twins. This range is indicated for the 86.8~K curve. (c) Schematic of
$\rho_{100}(\varepsilon_{\text{B1g}})$ for $T < T_\text{s}$; the underlying curve is not accessible for $-\varepsilon_\text{s} < \varepsilon_{\text{B1g}} < +\varepsilon_\text{s}$ due to the onset of
twinning, and the observed resistivity instead interpolates over this range. (d) Temperature ramps at three values of $\varepsilon_{\text{B1g}}$. (e) $T_\text{s}$ versus strain for low strains.  The
shaded regions indicate the transition width, defined by $d^2\rho_{100}/dT^2$ crossing half its maximum value. }

\newcommand{\figSampleBCompleteCaption}{(a) $\rho_{100}$ of Sample B over a wide temperature and strain range. Points are data from temperature ramps at constant $\varepsilon_\text{B1g}$, and lines
from strain ramps at constant temperature. $\varepsilon_\text{s}(T)$, taken as the data of Ref.~\cite{Kothapalli16NatComm} scaled in $T$ and $\varepsilon$ to match the data here, is indicated at each
temperature. (b) $\rho_{100}(T)$ for $\varepsilon_\text{B1g} = -0.25 \times 10^{-2}$, where the sample is fully detwinned at all $T$. (c) Close-up of the data in panel (a) at 14.6 and 36.9~K. The
squares mark points where the position along the $\varepsilon_\text{B1g}$ axis was adjusted to correct for plastic deformation of the platform; see Appendix section 4 for details.} 

\newcommand{\figMontgomeryCaption}{Data from Sample C, the Montgomery-configuration sample. (a) $\rho_{100}$ (left) and $\rho_{010}$ (right), from strain ramps at various fixed temperatures.  The
hysteresis is shown for two temperatures. The vertical ticks mark $-\varepsilon_\text{s}(T)$, taken from Ref.~\cite{Kothapalli16NatComm} and scaled in temperature to match the $T_\text{s}$ observed
here. (b) $\rho_{\text{B1g}} \equiv (\rho_{100} - \rho_{010})/2$, derived from the data in panel (a), at temperatures above $T_\text{s}$. Note that by symmetry $\rho_\text{B1g}(T>T_\text{s})=0$ at
$\varepsilon_\text{B1g}=0$, but measurement error gives a small deviation from this.  (c) $\rho_{\text{A1g}} \equiv (\rho_{100} + \rho_{010})/2$.}

\newcommand{\figResistivityAnisotropyCaption}{Nematic resistive anisotropy. (a) The spontaneous resistive anisotropy below $T_\text{s}$, obtained as described in the text. The inset shows $\rho_a$
and $\rho_b$ near $T_\text{s}$.  (b) B\textsubscript{1g} elastoresistivity, $(1/\rho_\text{A1g})d\rho_\text{B1g}/d\varepsilon_\text{B1g}$, obtained from both long strain ramps and from a small
oscillating strain. For Sample B, $\rho_{010}$ was not measured, so the $[100]$ elastoresistivity $(1/\rho_{100}) \times d\rho_{100}/d\varepsilon_{100}$ is plotted instead.  Fits are to a Curie-Weiss
form; see the text. (c) Resistivity anisotropy of Sample C against strain at $T \approx T_\text{s}$. The curve has been shifted vertically to set $\rho_\text{B1g} = 0$ at $\varepsilon_\text{B1g}=0$,
cancelling a small geometrical error in the measurement.  Also shown is an estimate of the strain-induced nematicity $\psi$, taken as the $xz$-$yz$ energy splitting at the $X$ point, obtained from
evaluation of Ginzburg-Landau parameters.}

\newcommand{\figBiaxialAndTcCaption}{Effect of biaxial strain. (a) A\textsubscript{1g} elastoresistivity $(1/\rho_\text{A1g})d\rho_\text{A1g}/d\varepsilon_\text{A1g}$ versus $T$ of Sample C,
determined as explained in the text. (b) $T_\text{c}$ versus strain, determined as the temperature where the resistivity crosses specific values, as shown in the inset.  Note that within the twinned
region, $\varepsilon_\text{B1g}$ does not couple locally to the sample, and instead the effect on $T_\text{c}$ is through the applied A\textsubscript{1g} component of the strain. When the platform
deformation is elastic, this is $\varepsilon_\text{A1g} = 0.52\varepsilon_\text{B1g}$. The observed slope therefore corresponds to $dT_\text{c}/d\varepsilon_\text{A1g} = -450$~K.}

\newcommand{\figSampleCTempRamps}{(a--b) Temperature ramp data from Sample C; panel (a) shows $\rho_{100}$ and panel (b) $\rho_{010}$. (c) Change in slope $d\rho/d\varepsilon_\text{B1g}$ across
$\varepsilon_\text{B1g} = 0$. This quantity is proportional to the twin boundary contribution to sample resistivity at $\varepsilon_\text{B1g}=0$.}

\newcommand{\figAppendixACaption}{Nematic resistivity anisotropy $(\rho_a - \rho_b)/(\rho_a + \rho_b)$ of Sample C, derived from the strain-ramp data shown in Fig.~\ref{figMontgomery}(a). This
determination is based on extraction of equilibrium slopes $d\rho/d\varepsilon_\text{B1g}$ at $\varepsilon_\text{B1g}=0$, obtained by averaging the observed increasing-$\varepsilon$ and
decreasing-$\varepsilon$ slopes at $\varepsilon_\text{B1g}=0$, as shown in the inset.}

\newcommand{\figAppendixBCaption}{$\rho_{110}$ versus $T$ of Sample A over a wide temperature range.}

\newcommand{\figPlasticCaption}{Plastic deformation of the platform. (a) $\rho_{100}$ of sample B versus displacement $D$ applied to the platform. The data sets were taken in the following order: (1)
Strain ramps at fixed temperature. (2) Temperature ramps at fixed strain. (3) Strain ramps at higher compression. The offset between data sets (1) and (3) is due to plastic deformation of the platform
that occurred over the course of the temperature ramps. (b) When data from set 3 are offset along the $D$ axis, the match with data set 1 is excellent.  (c) Schematic illustration of the process of
plastic deformation.  (d) Low-temperature resistivity measured before and after the platform plastic deformation. To compare data sets where $T_\text{c}$ was the same, the before data are taken at
$D = 0.8$~$\mu$m and the after data at $D = 1.6$~$\mu$m.}

\newcommand{\figTwinAnnealingCaption}{Annealing twin boundaries out of the sample by ramping the applied strain. See the Appendix text for details.}

\begin{document}

\title{The relationship between transport anisotropy and nematicity in FeSe}

\author{Jack Bartlett}
\thanks{These authors contributed equally.}
\affiliation{Max Planck Institute for Chemical Physics of Solids, N\"{o}thnitzer Str 40, 01187 Dresden, Germany}
\affiliation{SUPA, School of Physics and Astronomy, University of St. Andrews, St. Andrews KY16 9SS, United Kingdom}

\author{Alexander Steppke}
\thanks{These authors contributed equally.}
\email{steppke@cpfs.mpg.de}
\affiliation{Max Planck Institute for Chemical Physics of Solids, N\"{o}thnitzer Str 40, 01187 Dresden, Germany}


\author{Suguru Hosoi}
\affiliation{Department of Advanced Materials Science, University of Tokyo, Kashiwa, Chiba 277-8561, Japan}
\affiliation{Department of Materials Engineering Science, Osaka University, Toyonaka, Osaka 560-8531, Japan}

\author{Hilary Noad}
\affiliation{Max Planck Institute for Chemical Physics of Solids, N\"{o}thnitzer Str 40, 01187 Dresden, Germany}

\author{Joonbum Park}
\affiliation{Max Planck Institute for Chemical Physics of Solids, N\"{o}thnitzer Str 40, 01187 Dresden, Germany}

\author{Carsten Timm}
\affiliation{Institute of Theoretical Physics, Technische Universit\"{a}t Dresden, 01062 Dresden, Germany}
\affiliation{W\"{u}rzburg-Dresden Cluster of Excellence ct.qmat, Technische Universit\"{a}t Dresden, 01062 Dresden, Germany}

\author{Takasada Shibauchi}
\affiliation{Department of Advanced Materials Science, University of Tokyo, Kashiwa, Chiba 277-8561, Japan}

\author{Andrew P. Mackenzie}
\affiliation{Max Planck Institute for Chemical Physics of Solids, N\"{o}thnitzer Str 40, 01187 Dresden, Germany}
\affiliation{SUPA, School of Physics and Astronomy, University of St. Andrews, St. Andrews KY16 9SS, United Kingdom}

\author{Clifford W. Hicks}
\email{hicks@cpfs.mpg.de}
\affiliation{Max Planck Institute for Chemical Physics of Solids, N\"{o}thnitzer Str 40, 01187 Dresden, Germany}
\affiliation{School of Physics and Astronomy, University of Birmingham, Birmingham B15 2TT, U.K.}

\date{24 Nov 2020}

\begin{abstract}

The mechanism behind the nematicity of FeSe is not known. Through elastoresitivity measurements it has been shown to be an electronic instability. However, so far measurements have extended only
to small strains, where the response is linear. Here, we apply large elastic strains to FeSe, and perform two types of measurements. (1) Using applied strain to control twinning, the nematic resistive
anisotropy at temperatures below the nematic transition temperature $T_\text{s}$ is determined. (2) Resistive anisotropy is measured as nematicity is induced through applied strain at fixed
temperature above $T_\text{s}$. In both cases, as nematicity strengthens the resistive anisotropy peaks about about 7\%, then decreases. Below $\approx$40~K, the nematic resistive anisotropy changes
sign. We discuss possible implications of this behaviour for theories of nematicity. We report in addition: (1) Under experimentally accessible conditions with bulk crystals, stress, rather than
strain, is the conjugate field to the nematicity of FeSe. (2) At low temperatures the twin boundary resistance is $\sim$10\% of the sample resistance, and must be properly subtracted to extract
intrinsic resistivities. (3) Biaxial in-plane compression increases both in-plane resistivity and the superconducting critical temperature $T_\text{c}$, consistent with a strong role of the $yz$
orbital in the electronic correlations.

\end{abstract}

\maketitle

At an electronic-nematic transition, electronic interactions drive a spontaneous reduction in rotational symmetry without introducing translational or time-reversal symmetry breaking.  Electronic
nematicity affects all the Fermi surfaces of a metal, and therefore its fluctuations can have powerful effects~\cite{Licciardello19Nature, Lederer15PRL}. It is potentially an integral part of the
high-temperature superconductivity of iron-based and cuprate superconductors~\cite{Murayama19NatComm}, and the mechanisms behind it are therefore a topic of interest.

In many iron-based superconductors, nematicity occurs in close proximity to a transition into unidirectional spin density wave order, suggesting that it is a melted form of the magnetic
order~\cite{Fernandes12PRB, Fernandes14NatPhys}. In contrast, the nematic transition of FeSe occurs, at 92~K, without a subsequent magnetic transition. Whereas in other iron-based superconductors
magnetic and lattice fluctuations are linked by a scaling relationship, they are not so linked in FeSe~\cite{Fernandes13PRL, Boehmer15PRL, Baek14NatMat}. In spite of these differences, there are
similarities between FeSe and other iron-based superconductors that suggest that their nematicities are related. For example, unidirectional magnetic order can be induced in
FeSe~\cite{Terashima15JPSJ, Boehmer19PRB}, and the nematic electronic structure as observed in angle-resolved photoemission qualitatively matches that of BaFe$_2$As$_2$~\cite{Pfau19PRL, Yi19PRX}.
FeSe is a valuable reference material not only because of the absence of magnetic order, but also because of the absence of intrinsic dopant disorder, and the availability of high-quality,
vapor-transport-grown samples~\cite{Boehmer13PRB, Kasahara14PNAS}. 

Measurements of the strain dependence of resistivity, \textit{i.e.} the elastoresistivity, have shown that its nematicity, like that of other iron-based superconductors, is an electronic instability.
The key observation is that the resistive anisotropy $(\rho_{xx} - \rho_{yy})/(\rho_{xx} + \rho_{yy})$ varies with strain at a rate that diverges with cooling~\cite{Chu12Science, Kuo16Science,
Watson15PRB, Hosoi16PNAS, Tanatar16PRL}. The resistive anisotropy is understood to be proportional to an underlying electronic anisotropy that can be quantified by a nematic order parameter $\psi$. On
a clamped lattice, $\psi$ would transition to a nonzero value at a bare transition temperature $T_\text{s,0}$, but the elastic compliance of the lattice raises the transition temperature to
$T_\text{s} > T_\text{s,0}$. For $T>T_\text{s}$, applied anisotropic strain $\varepsilon$ induces nonzero $\psi$ through electron-lattice coupling, with a susceptibility $d\psi/d\varepsilon$ that
diverges (with divergence temperature $T_\text{s,0}$) as the sample is cooled. Therefore, because resistive anisotropy is proportional to $\psi$, its dependence on strain also steepens with
cooling.

An assumption of a linear relationship between $\psi$ and resistive anisotropy has become deeply enough embedded that resistive anisotropy is often employed as a measure of $\psi$. Here, we explore
elastoresistivity at large $|\psi|$, where the relationship becomes strongly nonlinear. FeSe is considered to be a Hund's metal, meaning that interorbital charge fluctuations are suppressed by Hund's
coupling~\cite{deMedici14PRL}. Strong evidence for the importance of orbital character is provided by the fact that the magnitude of the superconducting gap correlates closely with $yz$ orbital
weight~\cite{Sprau17Science, Rhodes18PRB, Shibauchi20JPSJ}. Many of the strain effects that we observe here are also consistent with a prominent role of the $yz$ orbital in electronic correlations,
and we discuss how our data may constitute a test of theories of the nematicity of FeSe.

Two types of measurement are presented. (1) Resistive anisotropy is measured as a function of strain-induced nematicity at constant temperature $T \sim T_\text{s}$. (2) Strain-tuning is employed to
control the twinning as samples are cooled, allowing measurement of the intrinsic resistive anisotropy at temperatures below $T_\text{s}$. Although the $T$-dependent nematic resistive anisotropy has
been reported previously for a few iron-based compounds~\cite{Chu10Science, Kuo12PRB, Tanatar16PRL, Blomberg12PRB, Blomberg13NatComm}, these previous measurements have relied upon assumptions that
twin boundary resistance is negligible, and/or that a sustained stress applied to detwin samples is weak enough not to substantially alter the electronic structure, even though the iron-based
superconductors are extremely sensitive to uniaxial stress~\cite{Kuo12PRB, Lu14Science}. With strain-tuning, samples can be held in a fully or partially detwinned state without sustained application
of external anisotropic stress.

This paper is organized as follows. We first present our setup and methods, and then define the key parameters for discussion of elastoresistivity. We then present results for application of
anisotropic strain with principal axes rotated by $45^\circ$ from the nematic axes, in other words where it constitutes a transverse field to the nematicity~\cite{Maharaj17PNAS}. Results are then
presented for strain aligned with the nematic axes, where the response is much stronger. Our main result, the spontaneous nematic resistive anisotropy for $T < T_\text{s}$, in comparison with that
induced by strain at $T > T_\text{s}$, is shown in Fig.~\ref{figResistivityAnisotropy}. 

For orientation, the electronic structure of FeSe above and far below $T_\text{s}$ is illustrated schematically in Fig.~\ref{figSetup}(a). We work with the 1-Fe unit cell, in which the Fe-Fe bond
directions, and the principal axes of the nematicity, are the $\langle 100 \rangle$ directions. In the corresponding Brillouin zone, there is a hole pocket at the $\Gamma$ point, and two electron
pockets, one at the $X$ and the other at the $Y$ point. In the nematic state, where the $a$ lattice parameter becomes larger than $b$, the pocket at $X$ distorts into a peanut-like shape elongated
along $k_x$, signatures of the $Y$ pocket disappear from spectroscopic probes~\cite{Sprau17Science, Terashima14PRB, Watson15PRB, Watson15PRL, Watson17NJP, Rhodes20PRB}, and the hole pocket becomes
elongated along the $k_y$ direction.

\section{Methods}

\begin{figure}[thb]
\includegraphics[width=\columnwidth]{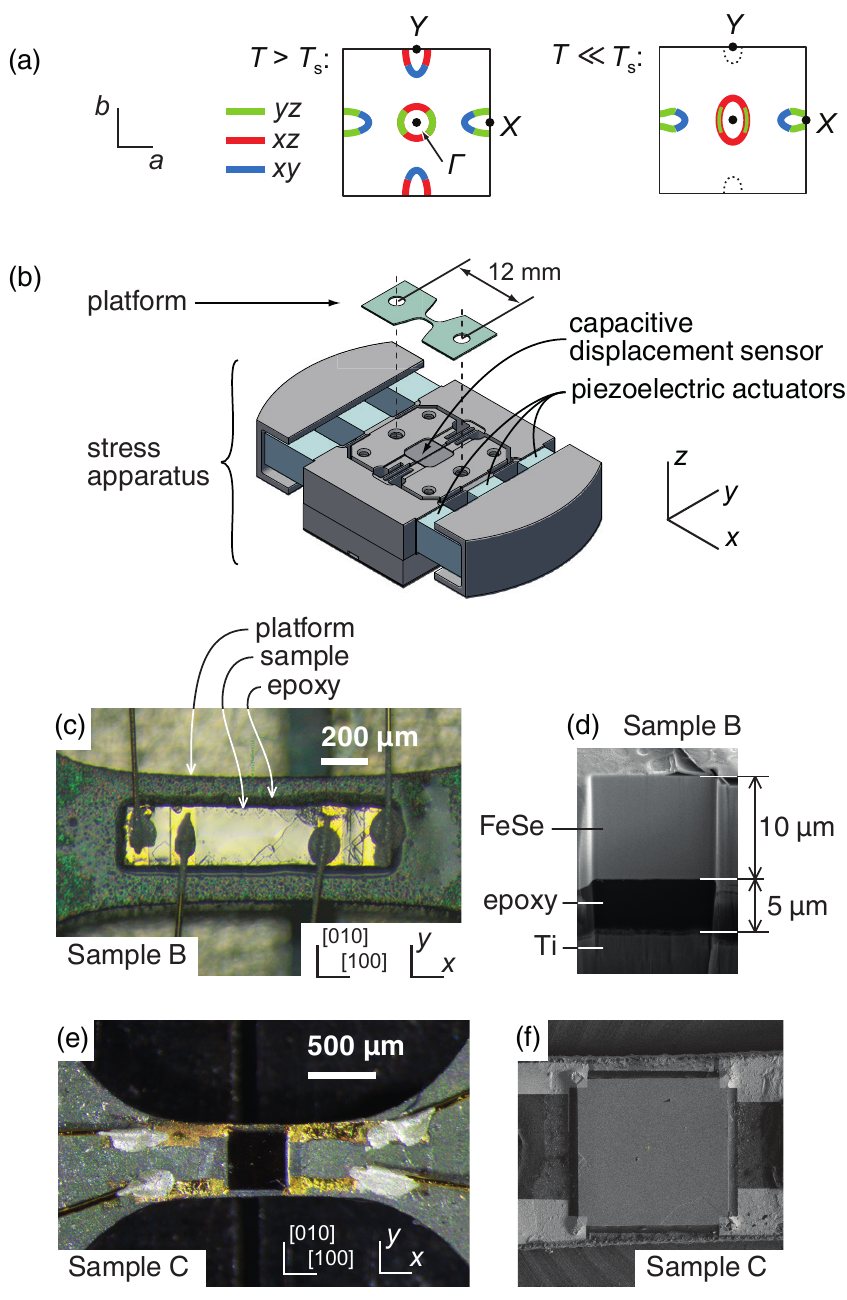}
\caption{\figSetupCaption{}}
\label{figSetup}
\end{figure}

To apply large strains to FeSe, we affix samples to platforms with a layer of epoxy (Masterbond\textsuperscript{\textregistered} EP29LPSP), and then apply stress to the platform; details of this
method are presented in Ref.~\cite{Park20RSI}. By preventing samples from buckling under compressive strain the platform allows samples to be very thin. This is helpful for FeSe
because it is a layered compound with a very low elastic limit for interlayer shear stress, which is minimized when samples are thin.  The epoxy that wicked up the sides of the sample may also have
served to hinder cleavage. A schematic of the setup is shown in Fig.~\ref{figSetup}(b), and images of mounted samples are shown in Fig.~\ref{figSetup}(c--f).

Here, the platforms are titanium sheets. The central portion is cut into a narrow neck within which stress is concentrated, and samples are attached to this neck. The platforms are then
mounted onto a piezoelectric-driven uniaxial stress apparatus. This apparatus incorporates a capacitive sensor of the applied displacement, and therefore of the longitudinal strain within the neck.

We report data from three samples. Sample A was cut for application of strain with $\langle 110 \rangle$ principal axes, and Samples B and C with $\langle 100 \rangle$ principal axes. Samples A and B
were prepared as shown in Fig.~1(c): bars with high length-to-width ratio, with contacts for measurement of resistivity along the sample length. Sample C, shown in Fig.~1(e--f), was prepared in a
Montgomery configuration for simultaneous measurement of longitudinal and transverse resistivities, as introduced for elastoresistivity measurements in Ref.~\cite{Kuo16Science}. The conversion from
measured resistances to longitudinal and transverse resistivities in the Montgomery geometry is discussed in Appendix section 1. In Appendix section~2, we consider the mechanics of strain transmission
from the platform to the sample, and show that the lengths and widths of the samples here are all long enough that to good precision both the longitudinal and transverse strains can be taken to be
locked to those in the platform.

Electrical contacts, fabricated from sputtered gold with no adhesion layer, were deposited on the samples' upper surfaces. The resistivity ratio $\rho_c / \rho_{ab}$ of FeSe appears not to have been
measured, however that of FeSe$_{0.4}$Te$_{0.6}$ is $\approx$70 at 15~K~\cite{Noji10JPSJ}. The length scale for current injected at the upper surface to spread out over the full sample thickness is $t
(\rho_{c}/\rho_{ab})^{1/2}$, where $t \sim 10$~$\mu$m is the sample thickness. This length scale is short enough that measurements here are not strongly affected by the $c$-axis resistivity.
For Sample C, the contacts also run down the sides of the sample.

\section{Key parameters}

\begin{table}[htbp]
        \centering
	\caption{Strain parameters. We take the 1-Fe unit cell, in which the $\langle 100 \rangle$ directions are Fe-Fe bond directions. Sample A is aligned so that stress is applied along the $[110]$
lattice direction; the strain along this axis, $\varepsilon_{110}$, is measured by the displacement sensor integrated into the stress cell. Samples B and C are aligned so that stress is applied along
the $[100]$ direction. $\nu = 0.32$ is the Poisson's ratio of the platform. The graphics illustrate the strain directions. We take the sign convention that $\varepsilon < 0$ denotes compression.}
        \begin{tabular}{|cc|}
                \hline
		\multicolumn{2}{|c|}{\textbf{Sample A}} \\
		\hline
		\rule{0pt}{5ex} \makecell{\includegraphics[width=12mm]{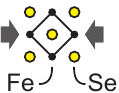}} & \makecell{$\varepsilon_\text{A1g} \equiv \frac{1}{2}(\varepsilon_{110} + \varepsilon_{1\bar{1}0}) = \frac{1}{2}(1 - \nu)\varepsilon_{110} = 0.34 \varepsilon_{110}$ \\[1ex] 
$\varepsilon_\text{B2g} \equiv \frac{1}{2}(\varepsilon_{110} - \varepsilon_{1\bar{1}0}) = \frac{1}{2}(1 + \nu)\varepsilon_{110} = 0.66\varepsilon_{110}$} \\[3ex]
		\hline
		\multicolumn{2}{|c|}{\textbf{Samples B and C}} \\
		\hline
		\rule{0pt}{5ex} \makecell{\includegraphics[width=12mm]{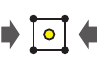}} & \makecell{$\varepsilon_\text{A1g} \equiv \frac{1}{2}(\varepsilon_{100} + \varepsilon_{010}) = \frac{1}{2}(1 - \nu)\varepsilon_{110} = 0.34\varepsilon_{100}$ \\[1ex] 
		$\varepsilon_\text{B1g} \equiv \frac{1}{2}(\varepsilon_{100} - \varepsilon_{010}) = \frac{1}{2}(1 + \nu)\varepsilon_{110} = 0.66 \varepsilon_{010}$} \\[3ex]
                \hline
        \end{tabular}
\end{table}

The applied strain can be resolved into symmetric and antisymmetric components, and throughout this work it will be important to resolve their separate effects. Here, we define quantities for
discussion.  For Sample A, stress is applied along the $[110]$ lattice direction; the displacement sensor in the stress cell measures the strain along this axis, $\varepsilon_{110}$. The transverse
strain $\varepsilon_{1\bar{1}0}$ is given by $\varepsilon_{1\bar{1}0} = -\nu \varepsilon_{110}$, where $\nu = 0.32$ is the Poisson's ratio of the platform. The symmetric component of the strain field
is $\varepsilon_\text{A1g} \equiv \frac{1}{2}(\varepsilon_{110} + \varepsilon_{1\bar{1}0})$, which comes to $0.34 \varepsilon_{110}$, while the antisymmetric component is $\varepsilon_\text{B2g}
\equiv \frac{1}{2}(\varepsilon_{110} - \varepsilon_{1\bar{1}0}) = 0.66 \varepsilon_{110}$. These parameters, along with equivalent parameters for Samples B and C, are summarized in Table~I.  We also
label resistivities by the measurement axis: $\rho_{100}$, for example, is the resistivity along the $[100]$ direction.  For Sample C both $\rho_{100}$ and $\rho_{010}$ are measured, and so symmetric
and antisymmetric resistivities can be defined: $\rho_{\text{A1g}} \equiv \frac{1}{2}(\rho_{100} + \rho_{010})$ and $\rho_{\text{B1g}} \equiv \frac{1}{2}(\rho_{100} - \rho_{010})$.

We note that specifying lattice distortions becomes more complicated when the lattice twins. We adopt here the convention that $[110]$ for Sample A, and $[100]$ for Samples B and C,
always refer to the direction along the length of the platform. When the sample twins, we use $a$ and $b$ to refer to the directions along which the in-plane lattice constant lengthens and shrinks,
respectively; in other words, the $a$ and $b$ axes are defined locally, and the $[100]$ and $[110]$ directions globally.

For all samples, the applied strain will also generate a $c$-axis strain in the sample, $\varepsilon_{001} = -2c_{13} \varepsilon_\text{A1g}/c_{33}$.  $c$-axis strain preserves the tetragonal symmetry
of the $T > T_\text{s}$ lattice, and therefore is in the A\textsubscript{1g} representation. When we discuss A\textsubscript{1g} strain it should be understood that it includes this associated
$c$-axis strain.

Because the aim of this work is to explore the nonlinear regime, we do not apply the elastoresistivity matrix formalism introduced in Ref.~\cite{Kuo13PRB}. For comparison with previous results we note
that the quantity $(1/\rho_\text{A1g})d\rho_\text{B1g}/d\varepsilon_\text{B1g}$ at $\varepsilon_{\text{B1g}}=0$ is equal to $m_{11} - m_{12}$ in that formalism. Most previous elastoresistivity results
have been reported using the 2-Fe unit cell, in which $m_{11} - m_{12}$ transforms to $2m_{66}$.

\begin{figure}[tbph]
	\includegraphics[width=\columnwidth]{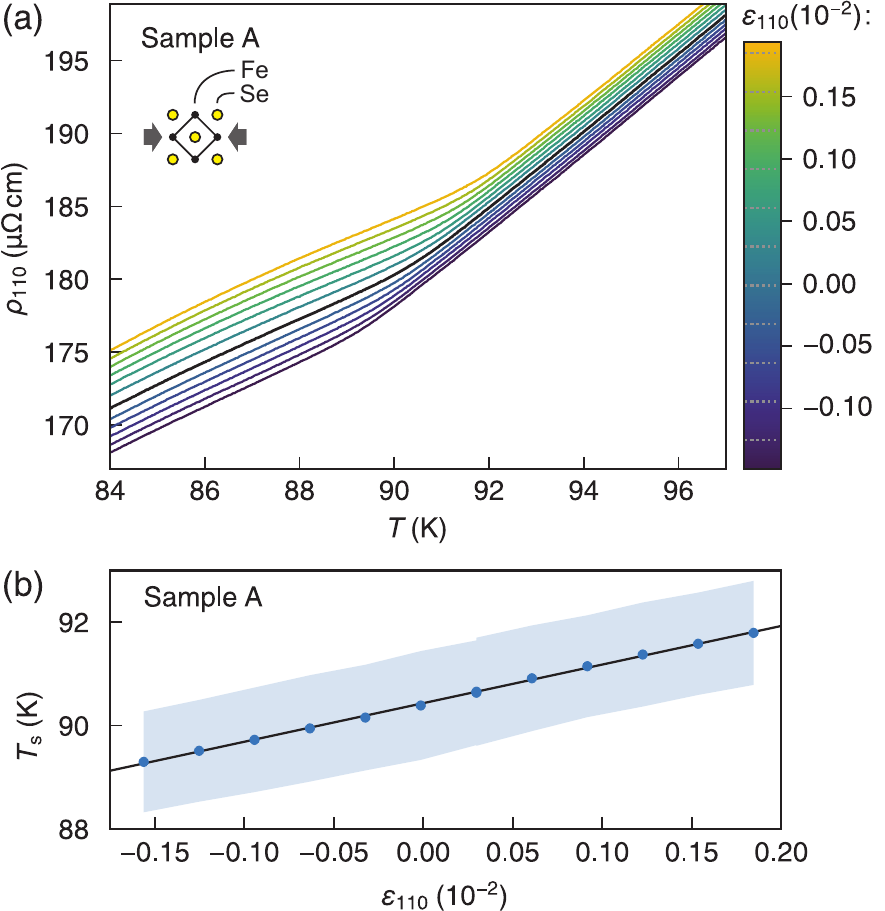}
	\caption{\figSampleACaption{}}
	\label{figSampleA}
\end{figure}

\section{Results: $\langle 110 \rangle$ strain}

Although strong transverse strain is predicted to enhance quantum fluctuations and suppress nematicity~\cite{Maharaj17PNAS, Ikeda18PRB}, the range of transverse strain explored here shifts
$T_\text{s}$ by only a few kelvin.  $T_\text{s}$ can be identified from an upturn in the resistivity, and, as shown in Fig.~\ref{figSampleA}, decreases at a modest rate with compression. Within our
strain range only a linear component of the strain dependence is resolved, with slope $dT_\text{s}/d\varepsilon_{110} = 750$~K. This slope is due to the $\text{A}_\text{1g}$ component of the applied
strain: under the tetragonal symmetry of FeSe at $T>T_\text{s}$, reversal of the sign of $\varepsilon_{\text{B2g}}$ gives a symmetrically equivalent strain, so coupling to $\varepsilon_{\text{B2g}}$
can give only strain-even components in the strain dependence of $T_\text{s}$.

$\varepsilon_\text{A1g} = 0.34\varepsilon_{110}$, so $dT_\text{s}/d\varepsilon_{110} = 750$~K corresponds to $dT_\text{s}/d\varepsilon_\text{A1g} = 2200$~K.  In Ref.~\cite{Kaluarachchi16PRB},
$T_\text{s}$ is found to be suppressed by compressive hydrostatic stress with an initial slope of 39~K/GPa. Using the elastic moduli of Ref.~\cite{Zvyagina13EPL}, this converts to
$dT_\text{s}/d\varepsilon_\text{A1g} \approx 6200$~K. (See Appendix section 3 for details.) The difference between this and our result allows, in principle, separation of the effect of $c$-axis strain
$\varepsilon_{001}$ and that of ``pure'' in-plane biaxial strain $\varepsilon_\text{A1g, pure}$ that has no associated $c$-axis strain. Applying again the elastic moduli from
Ref.~\cite{Zvyagina13EPL}, under in-plane uniaxial stress $\varepsilon_{001} = -0.3 \times \varepsilon_\text{A1g, pure}$, and under hydrostatic stress $\varepsilon_{001} = 1.0 \times
\varepsilon_\text{A1g, pure}$, so $\Delta T_\text{s} \approx (3200~\text{K}) \times \varepsilon_\text{A1g, pure} + (1000~\text{K}) \times \varepsilon_{001}$.

\section{Results: $\langle 100 \rangle$ strain}

\subsection{Stress-temperature versus strain-temperature phase diagram.} 

The effect of strain applied along the principal axes of the nematicity is much more dramatic. Before showing results, we discuss the differences between stress- and strain-temperature phase diagrams
for a nematic transition. The distinction between stress and strain is equivalent to that between magnetic field $H$ and magnetic induction $B$. When a ferromagnet is cooled through its Curie
temperature under nonzero $H$ the transition broadens into a crossover. Experimentally, controlled $H$ is applied by preparing samples to have a low demagnetization factor: thin bars parallel to the
applied field. In the opposite limit, of a thin plate perpendicular to the applied field, it is $B$ that is held fixed, and if $B/\mu_0$ is less than the spontaneous magnetization $M$ of the sample
then in general magnetic domains will form such that the sample's average magnetization matches the applied $B$. Domain formation under nonzero applied $B$ requires reversal of local magnetization, so
it is a first-order transition rather than a crossover.

\begin{figure}
	\includegraphics[width=\columnwidth]{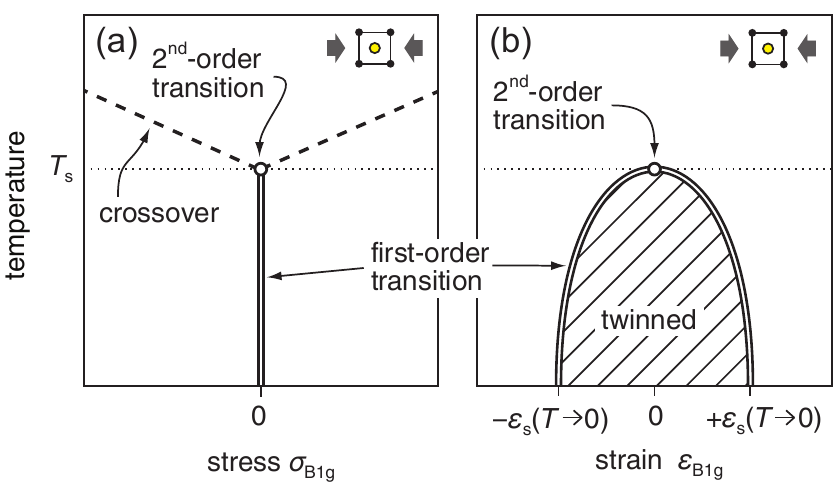}
	\caption{\figPhaseDiagramsCaption{}}
	\label{figPhaseDiagrams}
\end{figure}

For nematic compounds, the difference between stress and strain is illustrated in Fig.~\ref{figPhaseDiagrams}.  In the stress-temperature phase diagram, a first-order transition line corresponding to
reversal of the nematicity runs along the zero-stress axis from $T = T_\text{s}$ to $T \rightarrow 0$.  In the strain-temperature phase diagram, on the other hand, there are two lines of first-order
transitions.  The structural distortion in FeSe is to high precision a B\textsubscript{1g} distortion, meaning that $b$ contracts by nearly the same amount as $a$ lengthens~\cite{Kothapalli16NatComm,
Margadonna08ChemComm, McQueen09PRL}. Therefore, the nematicity-induced structural distortion can be described as a spontaneous local strain $\varepsilon_\text{B1g, local} = \pm
\varepsilon_\text{s}(T)$, where the quantity $\varepsilon_\text{s}$ is termed the structural strain. The average strain in the sample must match that of the platform, but when
$|\varepsilon_\text{B1g}| < \varepsilon_\text{s}(T)$ twin formation is favored, and the applied strain sets the equilibrium twin volume ratio. Like formation of magnetic domains under nonzero $B$,
formation of twinned domains under nonzero applied $\varepsilon_{\text{B1g}}$ is a first-order process, so the twinned region is bounded by first-order transitions.

In the stress-temperature phase diagram there will be resolvable crossover lines at $T > T_\text{s}$: when the applied stress is small, there will be a small temperature range over which the
nematicity-driven strain increases at a rapid but non-divergent rate. In this sense, stress acts as a classic conjugate field. We present some evidence below on whether equivalent crossover lines are
discernable in the strain-temperature phase diagram.

\begin{figure}[tbhp]
	\includegraphics[width=\columnwidth]{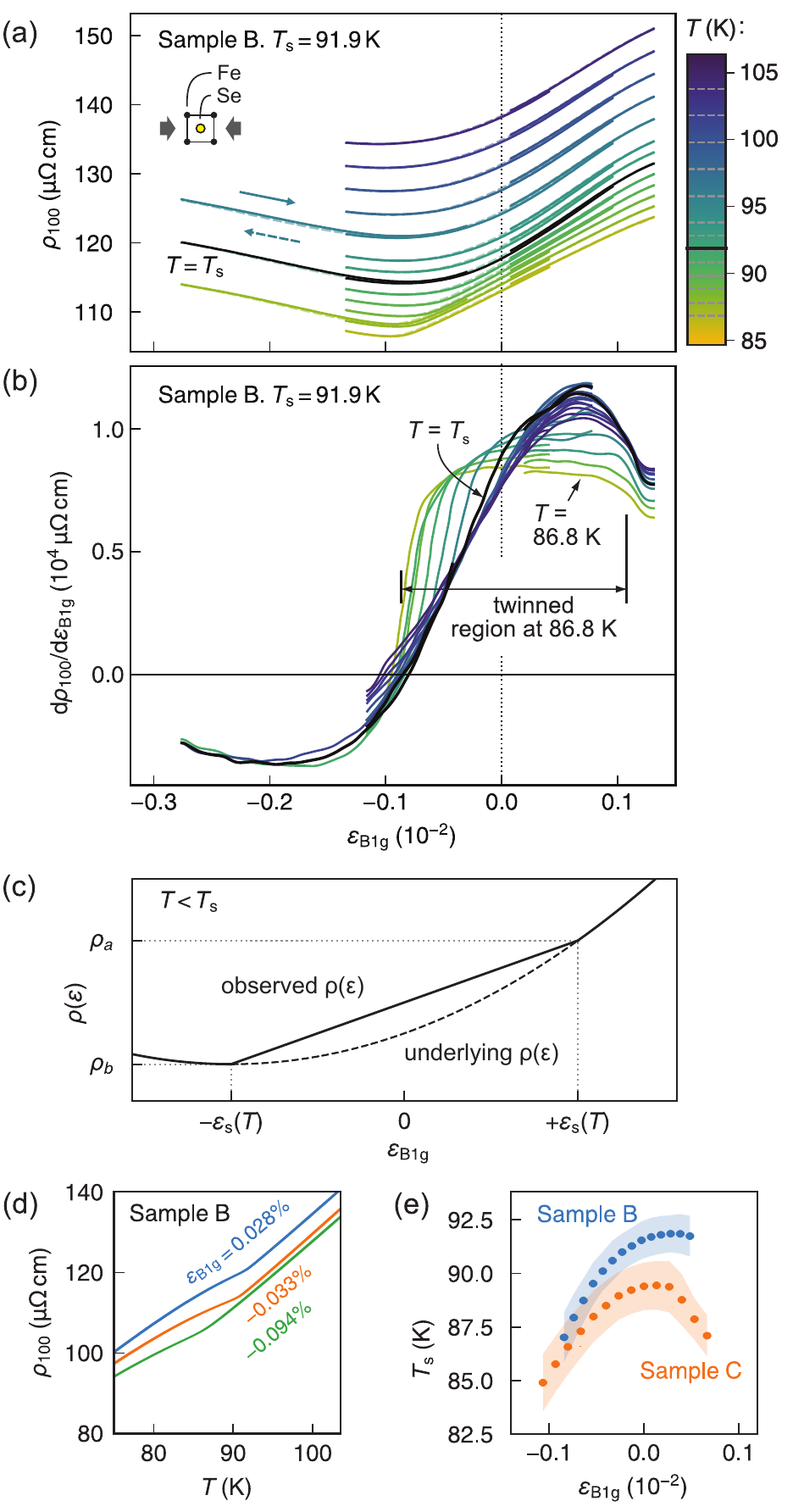}
	\caption{\figNearTsCaption{}}
	\label{figNearTs}
\end{figure}

\subsection{Sample B, \boldmath{$T \sim T_\text{s}$}}

Measurements of resistivity confirm this qualitative form of strain-temperature phase diagram. To facilitate comparison with measurements of $\varepsilon_\text{s}$, we now plot data against the
antisymmetric strain $\varepsilon_\text{B1g}$. $\rho_{100}(\varepsilon_{\text{B1g}})$ of Sample B for $T \sim T_\text{s}$ is shown in Fig.~\ref{figNearTs}(a), and the derivative
$d\rho_{100}/d\varepsilon_{\text{B1g}}$ in panel (b). The neutral strain point $\varepsilon_{\text{B1g}}=0$ is determined as the strain where the twin boundary density for $T<T_\text{s}$ is highest;
these data are shown below. Above $T_\text{s}$, the strain dependence of $\rho_{100}$ is seen to have substantial nonlinearity even over a relatively small strain range $|\varepsilon_\text{B1g}| < 0.1
\cdot 10^{-2}$. Its slope is largest near, though not precisely at, $\varepsilon_\text{B1g}=0$. 

As $T$ is reduced below $T_\text{s}$, the onset of twinning changes the form of $\rho_{100}(\varepsilon_{\text{B1g}})$: a range of strain appears over which
$d\rho_{100}/d\varepsilon_{\text{B1g}}$ becomes nearly constant. This change is easiest to see in Fig.~\ref{figNearTs}(b), where we have marked the twinned region for the 86.8~K curve. The origin of
this behavior is illustrated schematically in Fig.~\ref{figNearTs}(c). Within each twin domain the resistivities along the local $a$ and $b$ axes are $\rho_a$ and $\rho_b$, and the equilibrium
twin volume ratio is a linear function of applied strain. Therefore, the observed bulk resistivity is an interpolation between $\rho_b$ at $\varepsilon_{\text{B1g}} = -\varepsilon_\text{s}$ and
$\rho_a$ at $\varepsilon_{\text{B1g}}= +\varepsilon_\text{s}$, that to high precision is linear under two conditions that are both satisfied here. (1) $|(\rho_a - \rho_b)/(\rho_a +
\rho_b)|$ is much less than 1, so that redistribution of current into lower-resistivity domains does not substantially alter the observed bulk resistivity. (2) The domain wall resistance is
negligible, which we show later to be the case for $T$ near $T_\text{s}$. 

Even though the transitions into the twinned region must, when $\varepsilon_\text{B1g} \neq 0$, be first-order, no hysteresis is resolved, indicating that the energy barrier for twin formation is low.
Separately, close inspection of Figs.~\ref{figNearTs}(b) reveals that twinning does not initially onset right at $\varepsilon_\text{B1g}=0$, but slightly on the tensile side.  This asymmetry is due to
the A\textsubscript{1g} component of the applied strain: as shown with Sample A in Fig.~\ref{figSampleA}, tensile A\textsubscript{1g} strain increases $T_\text{s}$.

$\rho$ versus temperature at a few nonzero $\varepsilon_{\text{B1g}}$ are shown in Fig.~\ref{figNearTs}(d), and Fig.~\ref{figNearTs}(e) shows $T_\text{s}$ derived from such temperature sweeps as a
function of strain. For both Samples B and C, $T_\text{s}$ follows a downward quadratic form, consistent with the schematic strain-temperature phase diagram illustrated in
Fig.~\ref{figPhaseDiagrams}(b).  

\subsection{Sample B, \boldmath{$T < T_\text{s}$}}

\begin{figure}[thbp]
	\includegraphics[width=0.99\linewidth]{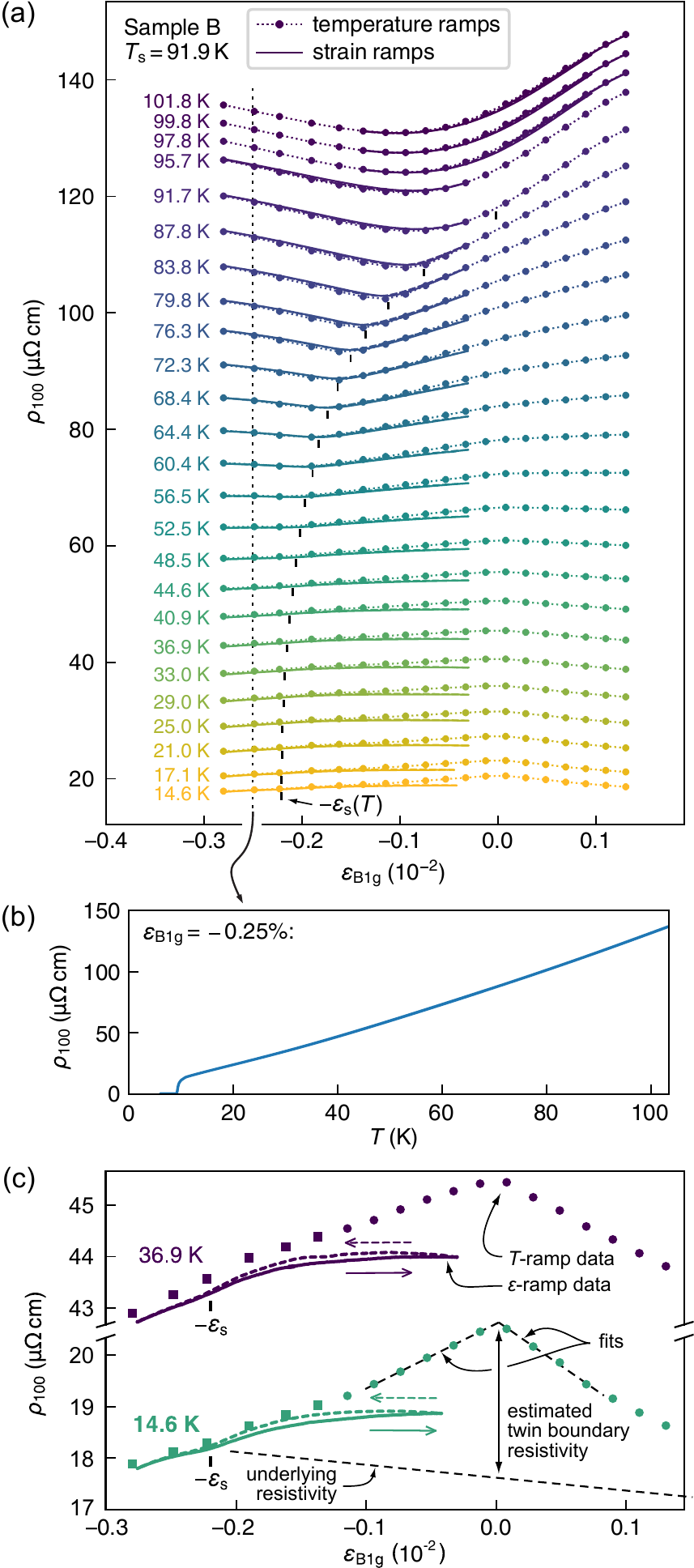}
	\caption{\figSampleBCompleteCaption{}}
	\label{figSampleBComplete}
\end{figure}

Fig.~\ref{figSampleBComplete}(a) shows $\rho_{100}$ of Sample B over a much wider temperature and strain range. Here, the contribution of twin boundaries to the total sample resistance becomes
apparent.  Two data sets are shown: strain ramps in which $T$ was incremented at $\varepsilon_\text{B1g} < - \varepsilon_\text{s}(T)$, and temperature ramps in which strain was incremented at $T >
T_\text{s}$. The maximum compression reached was $\varepsilon_\text{B1g} = -0.28 \times 10^{-2}$, which exceeds the spontaneous $T \rightarrow 0$ structural distortion of FeSe and fully detwins the
sample at all temperatures.  It corresponds to a longitudinal strain of $\varepsilon_{100} = -0.42 \times 10^{-2}$, and was large enough to exceed the elastic limit of the platform. Plastic
deformation of the platform introduced an anomalous offset between $\varepsilon_\text{B1g}$ and $\varepsilon_\text{A1g}$ at large strains. Data shown in Appendix section 4, where the plastic
deformation is described in more detail, show that the resistivity of FeSe depends much more sensitively on $\varepsilon_\text{B1g}$ than $\varepsilon_\text{A1g}$, and so we continue to plot data
against $\varepsilon_\text{B1g}$.  Crucially, the sample residual resistivity did not change, showing that its own deformation remained elastic even as the platform deformed plastically.

For $T$ above $\approx 60$~K, the structural strain $\varepsilon_\text{s}(T)$ can be identified by a sharp change in slope $d\rho_{100}/d\varepsilon_\text{B1g}$, as seen also in
Figs.~\ref{figNearTs}(a--b). To obtain $\varepsilon_\text{s}$ at all temperatures, we scale $\varepsilon_\text{s}(T)$ from the X-ray diffraction data of Ref.~\cite{Kothapalli16NatComm} in temperature
to match $T_\text{s}$ of this sample, and in strain to match the locations of the cusps. This procedure gives $\varepsilon_\text{s}(T \rightarrow 0) = 0.22 \cdot 10^{-2}$.  For comparison,
$\varepsilon_\text{s}(T \rightarrow 0) = 0.27 \times 10^{-2}$ and $0.23 \times 10^{-2}$ were obtained respectively in Refs.~\cite{Kothapalli16NatComm} and~\cite{Frandsen19PRB} by X-ray diffraction,
$0.24 \times 10^{-2}$ and $0.25 \times 10^{-2}$ in Refs.~\cite{Wang16NatMat} and~\cite{Rahn15PRB} by neutron scattering, and $0.22 \times 10^{-2}$ in Ref.~\cite{Boehmer13PRB} by dilatometry
measurements.

Fig.~\ref{figSampleBComplete}(b) shows $\rho_{100}(T)$ at fixed strain $\varepsilon_\text{B1g} = -0.25 \times 10^{-2}$, where the sample is detwinned at all temperatures.  $\rho_{100}$ evolves
smoothly from $T_\text{c}$ to above $T_\text{s}$, with no feature apparent that could be identified as a nematic crossover. In other words, it does not appear to be useful to consider strain as a
conjugate field to nematicity in FeSe, because even under a strain that is only barely large enough to detwin the sample any nematic crossover appears to be so broad as to be indistinguishable from
the background.

We now discuss twin boundaries.  For $|\varepsilon_\text{B1g}| < \varepsilon_\text{s}(T)$, $\rho_{100}$ from the temperature ramps systematically exceeds that from the strain ramps.  Panel (c) shows a
closeup of data at 36.9 and 14.6 K: the $T$-ramp data have a peaked form that the strain-ramp data do not. The magnitude of this peak is very similar at the two temperatures, even though the intrinsic
resistivity at 36.9 K is more than double that at 14.6 K, which shows that its origin is extrinsic. It is due to twin boundaries. The elastic mismatch between the sample, which distorts
orthorhombically, and the platform, which does not, will be strongest at $\varepsilon_\text{B1g}=0$, leading to a peak in the equilibrium twin boundary density. This peak is resolvable for
temperatures up to $\sim$70~K, at a temperature-independent strain, which we therefore identify as the neutral strain point $\varepsilon_\text{B1g}=0$. Evidence for twinning is also directly visible
in the strain-ramp data in Fig.~\ref{figSampleBComplete}(c), there is hysteresis for $|\varepsilon_\text{B1g}| < \varepsilon_\text{s}$ that closes when $|\varepsilon_\text{B1g}| >
\varepsilon_\text{s}$. In Appendix section 5 we show that ramping the strain back and forth can partially anneal twin boundaries out of the sample.

A method to estimate the twin boundary contribution to the measured resistivity is illustrated in Fig.~\ref{figSampleBComplete}(c). For a B\textsubscript{1g} lattice distortion, the twin boundary
density is expected to be symmetric about $\varepsilon_\text{B1g} = 0$. Furthermore, because twin boundaries are oriented along $\langle 110 \rangle$ directions~\cite{Tanatar16PRL}, no average change
in twin boundary orientation is expected for strain with $\langle 100 \rangle$ principal axes.  We therefore fit lines to the temperature-ramp data on either side of the cusp and average their slopes
to obtain an underlying slope, meaning the slope $d\rho_{100}/d\varepsilon_\text{B1g}$ that would be observed if the twin boundary resistance were zero. The line labelled ``underlying resistivity'' in
Fig.~\ref{figSampleBComplete}(c) is a line of this slope placed to intersect the data at $\varepsilon_\text{B1g} = -\varepsilon_\text{s}$, where the sample is de-twinned. In this way, we find that at
14.6~K the twin boundary contribution to the sample resistance is as high as 15\%, for this sample geometry. Twin boundary density may be lower for thicker and/or free-standing samples.

\begin{figure}[thbp]
\includegraphics[width=0.99\linewidth]{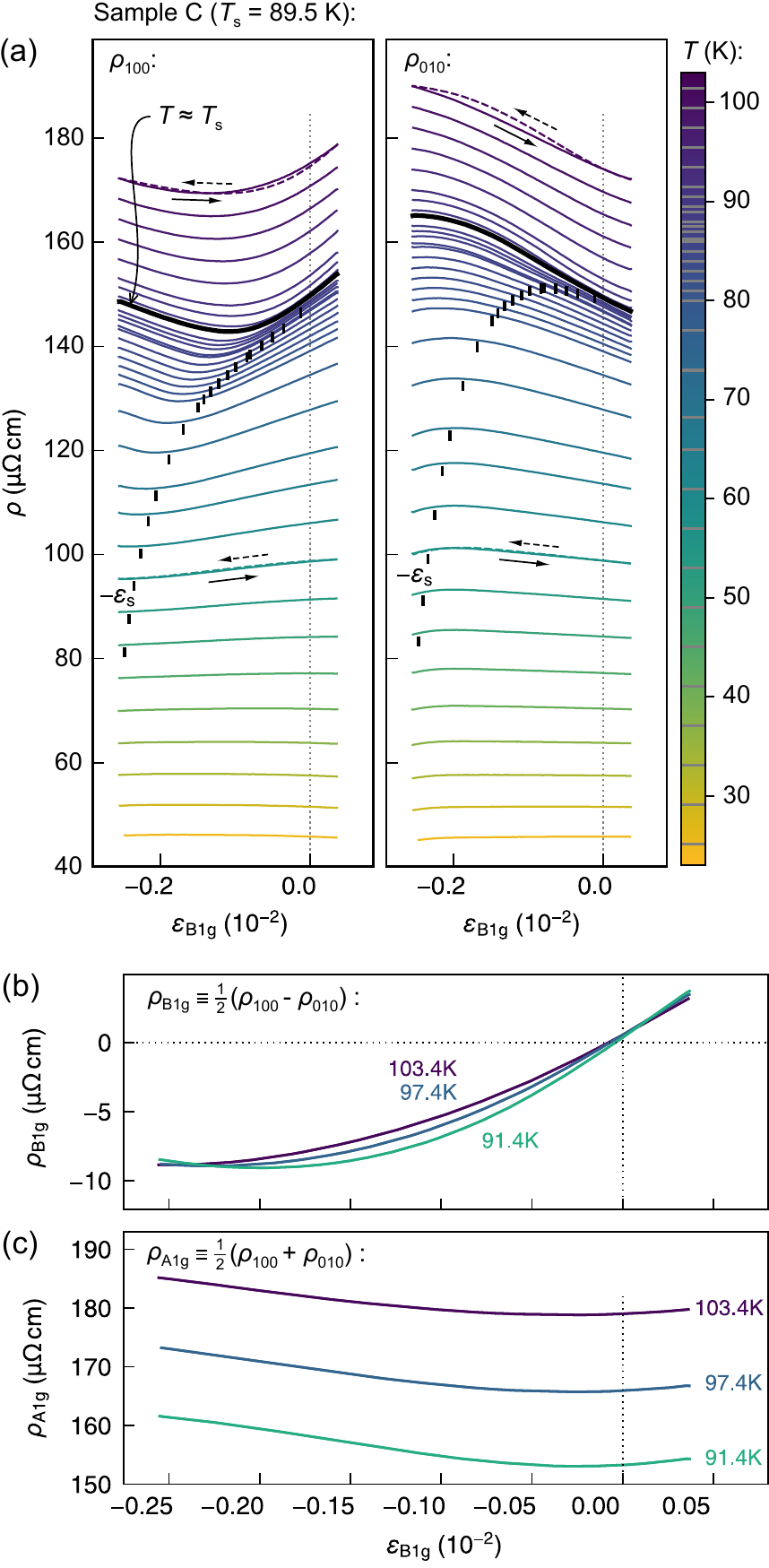}
\caption{\figMontgomeryCaption{}}
\label{figMontgomery}
\end{figure}

\subsection{Sample C}

In Sample C both the longitudinal and transverse resistivities, $\rho_{100}$ and $\rho_{010}$, were measured. Results from strain ramps are shown in Fig.~\ref{figMontgomery}, and from $T$ ramps in
Appendix section 6. The neutral strain point $\varepsilon_\text{B1g}=0$ was again taken as the strain where twin boundary density in the $T$-ramp data was highest. Around $\varepsilon_{\text{B1g}}=0$
and at temperatures near $T_\text{s}$, $\rho_{100}$ and $\rho_{010}$ vary strongly and oppositely with $\varepsilon_{\text{B1g}}$, confirming previous reports that the low-strain elastoresistivity of
FeSe is dominantly in the B\textsubscript{1g} channel~\cite{Hosoi16PNAS, Tanatar16PRL}. Below $T_\text{s}$, the twinning transition at $\varepsilon_\text{B1g} = -\varepsilon_\text{s}(T)$ is broader
than for Sample B. Although this could indicate lower sample quality, we also note that strain inhomogeneity will generally be worse in a square sample geometry than in the linear geometry of
Sample B. To estimate $\varepsilon_\text{s}(T)$ for Sample C, we scale $\varepsilon_\text{s}(T)$ reported in Ref.~\cite{Kothapalli16NatComm} in temperature to match the observed $T_\text{s}$ of Sample
C, but we do not scale it in strain.  

The antisymmetric resistivity $\rho_{\text{B1g}}$ for temperatures near $T_\text{s}$ is plotted in panel (b). Here it can be seen that although $|\rho_\text{B1g}|$ initially grows rapidly with
strain-induced nematicity, it eventually reaches a maximum; just above $T_\text{s}$, this occurs at $\varepsilon_\text{B1g} \approx -0.19 \cdot 10^{-2}$. The symmetric resistivity $\rho_\text{A1g}$
is plotted in panel (c). For $T \gtrsim T_\text{s}$, $\rho_\text{A1g}$ is a minimum near $\varepsilon_\text{B1g} = 0$, and as $T$ is reduced towards $T_\text{s}$ this minimum becomes sharper.

There are indications that other iron-based superconductors will have similar behavior. In Ba(Fe$_{0.975}$Co$_{0.025}$)$_2$As$_2$ (for which the nematicity also aligns with the $\langle 100 \rangle$
directions)
$\rho_{100}$ and $\rho_{010}$ both have upward curvature against $\varepsilon_\text{B1g}$, that grows sharper as $T$ is reduced to $T \approx
T_\text{s}$~\cite{Palmstrom17PRB}, suggesting that in this material too $\rho_\text{A1g}$ is a minimum for $\varepsilon_\text{B1g} \approx 0$.  For BaFe$_2$As$_2$, $\rho_{100}$ near $T_\text{s}$ has
been observed to have an S-shaped dependence on $\varepsilon_{100}$, with the steepest slope appearing near $\varepsilon_{100} = 0$~\cite{Liu16PRL}, matching the qualitative form (though with opposite
sign) of $\rho_{100}(\varepsilon_{\text{B1g}})$ observed here. Similar behavior is seen in Sr$_{1-x}$Ba$_x$Fe$_{1.97}$Ni$_{0.03}$As$_2$~\cite{Mao18CPB}.

\section{Effect of biaxial strain.}

Data from Sample C allow effects of the A\textsubscript{1g} and B\textsubscript{1g} strain components to be separated.  The A\textsubscript{1g} elastoresistivity
$d\rho_\text{A1g}/d\varepsilon_\text{A1g}$ can be obtained by noting that within the twinned region B\textsubscript{1g} strain does not couple locally to the sample, because the local
B\textsubscript{1g} strain is fixed at $\pm \varepsilon_\text{s}(T)$, but A\textsubscript{1g} strain does couple locally. We take $\rho_\text{A1g}$ within the twinned region as $\rho_\text{A1g} =
(\rho_a + \rho_b)/2$, and now determine $d\rho_\text{A1g}/d\varepsilon_\text{A1g}$ at $\varepsilon_\text{B1g} = \varepsilon_\text{A1g} = 0$. 

Under the approximation of linear interpolation between $\rho_a$ and $\rho_b$ and neglecting twin boundary resistance,
$\rho_{100}$ and $\rho_{010}$ in the twinned region are given by
\begin{eqnarray}
\rho_{100} & = & f \rho_a + (1-f) \rho_b , \\
\rho_{010} & = & f \rho_b + (1-f) \rho_a ,
\end{eqnarray}
where $f = (\varepsilon_\text{s} + \varepsilon_\text{B1g})/2\varepsilon_\text{s}$ is the volume fraction of the sample with the nematic $a$ axis oriented along the long axis of the platform.
Differentiating with respect to $\varepsilon_\text{B1g}$ gives:
\begin{equation}
\frac{d\rho_{100}}{d\varepsilon_\text{B1g}}  = \frac{\rho_a - \rho_b}{2\varepsilon_\text{s}} + f\frac{d\rho_a}{d\varepsilon_\text{B1g}} + (1-f)\frac{d\rho_b}{d\varepsilon_\text{B1g}},
\end{equation}
\begin{equation}
\frac{d\rho_{010}}{d\varepsilon_\text{B1g}}  = \frac{\rho_b - \rho_a}{2\varepsilon_\text{s}} + f\frac{d\rho_b}{d\varepsilon_\text{B1g}} + (1-f)\frac{d\rho_a}{d\varepsilon_\text{B1g}}.
\end{equation}
Under the experimental conditions here, $d/d\varepsilon_\text{B1g} = (d\varepsilon_\text{A1g}/d\varepsilon_\text{B1g}) d/d\varepsilon_\text{A1g} = [(1-\nu)/(1+\nu)]d/d\varepsilon_\text{A1g}$. Summing
Eqs. (3) and (4) yields the A\textsubscript{1g} elastoresistivity:
\begin{equation}
\frac{d\rho_\text{A1g}}{d\varepsilon_\text{A1g}} = \frac{1+\nu}{2(1-\nu)} \left( \frac{d\rho_{100}}{d\varepsilon_\text{B1g}} + \frac{d\rho_{010}}{d\varepsilon_\text{B1g}} \right)
\end{equation}
To obtain underlying slopes $d\rho_{100}/d\varepsilon_\text{B1g}$ and $d\rho_{010}/d\varepsilon_\text{B1g}$, that is, that exclude the effect of twin boundaries, we average the observed slopes on
either side of $\varepsilon_\text{B1g}=0$, as shown in Fig.~\ref{figSampleBComplete}(c). 

\begin{figure}[htbp]
	\includegraphics[width=0.99\linewidth]{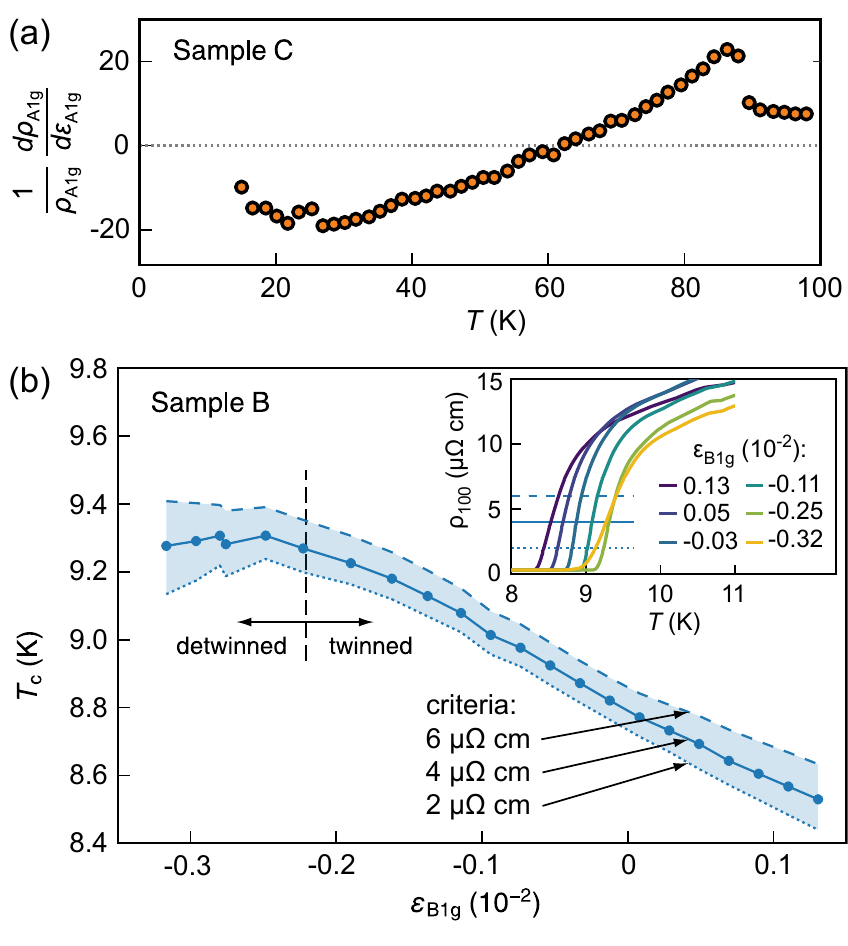}
	\caption{\figBiaxialAndTcCaption{}}
	\label{figBiaxialAndTc}
\end{figure}

The A\textsubscript{1g} elastoresistivity is shown in Fig.~\ref{figBiaxialAndTc}(a). It is normalized by $\rho_\text{A1g}$ at $\varepsilon_\text{B1g}=0$ with an estimate of the twin boundary
resistance subtracted (see Appendix section 6 for details). For temperatures below $\approx$60~K, $d\rho_\text{A1g}/d\varepsilon_\text{A1g} < 0$, meaning that biaxial compression increases the
average in-plane resistivity of FeSe. A similar temperature dependence is seen in the elastoresistivity of Sample A; see Appendix section 7.

We show in panel (b), with data from Sample B, that biaxial compression also increases $T_\text{c}$--- again, when the sample is twinned only the A\textsubscript{1g} component of the strain couples
locally. Both the increase in $T_\text{c}$ and $\rho_\text{A1g}$ are opposite to the generic expectation that compression should increase bandwidths. A similar correlation between resistivity and
$T_\text{c}$ is also seen in strained Sr$_2$RuO$_4$~\cite{Barber18PRL}.

At large $|\varepsilon_\text{B1g}|$, the plastic deformation of the platform causes a gradual relaxation of the applied A\textsubscript{1g} strain, and so for $\varepsilon_\text{B1g} \lesssim -0.15
\times 10^{-2}$ the $T_\text{c}$ curve bends downward subtly. For $\varepsilon_\text{B1g} < - \varepsilon_\text{s}$, the sample detwins, and the B\textsubscript{1g} component of the applied strain
couples locally to the sample. $T_\text{c}$ turns downward more sharply. Depending on the resistivity level selected as the criterion for $T_\text{c}$, it may even decrease. This behavior suggests
that increasing the lattice orthorhombicity is detrimental to superconductivity.

\section{The nematic resistive anisotropy}

\begin{figure}[htbp]
	\includegraphics[width=0.99\linewidth]{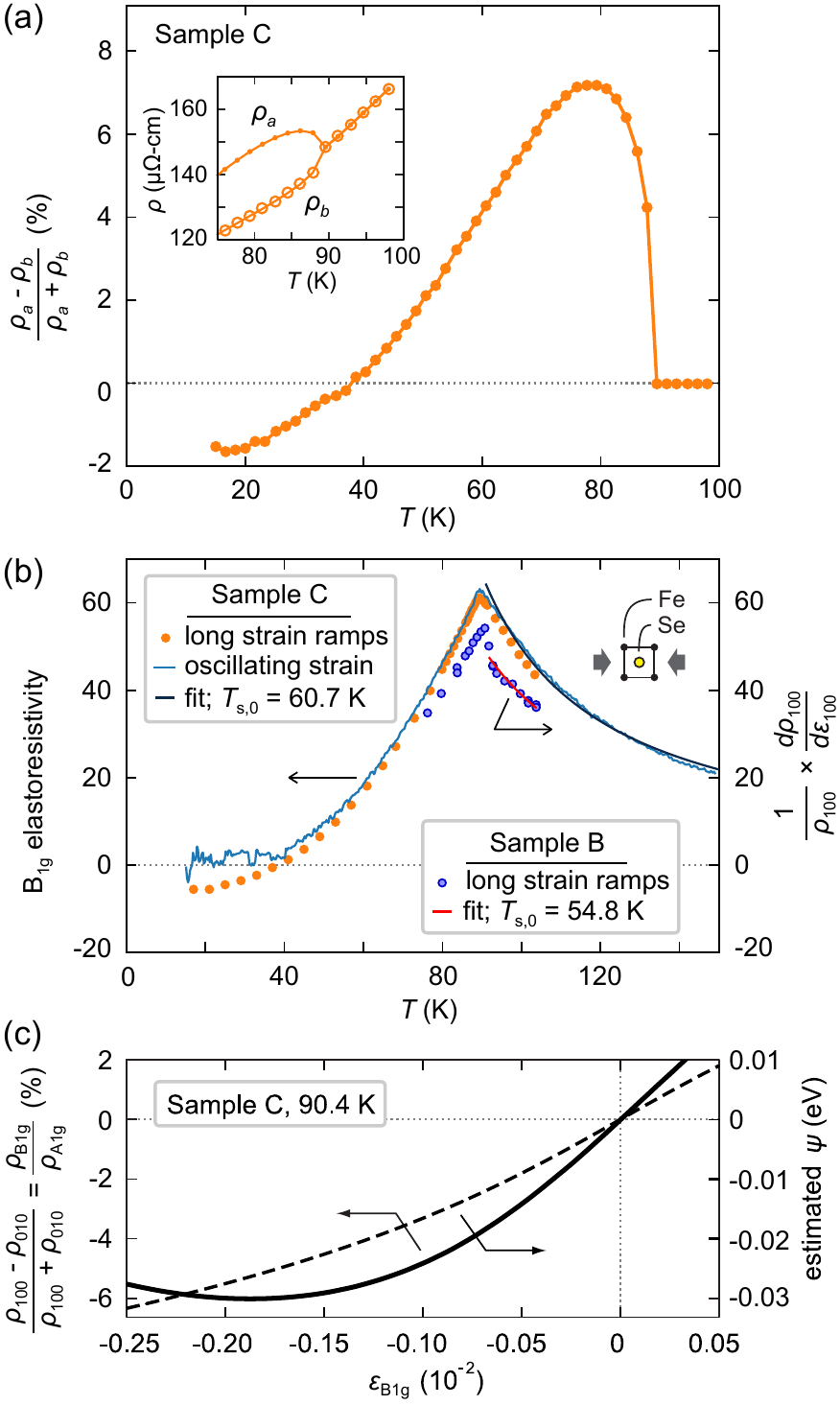}
	\caption{\figResistivityAnisotropyCaption{}}
	\label{figResistivityAnisotropy}
\end{figure}

We now report the central result of this paper, the nematic resistive anisotropy, both the spontaneous anisotropy below $T_\text{s}$ and that induced by strain at $T \sim T_\text{s}$.  We obtain
$\rho_a - \rho_b$ at $T<T_\text{s}$ by analyzing temperature-ramp data at small strains. At $\varepsilon_\text{B1g}=0$, $f$ in Eqs.~(1) and (2) is 0.5, yielding
\begin{equation}
\rho_a - \rho_b = \varepsilon_\text{s} \left( \frac{d\rho_{100}}{d\varepsilon_\text{B1g}} -
\frac{d\rho_{010}}{d\varepsilon_\text{B1g}}\right).
\end{equation}
The underlying slopes $d\rho_{100}/d\varepsilon_\text{B1g}$ and $d\rho_{010}/d\varepsilon_\text{B1g}$ are obtained, as before, by averaging the observed slopes from $\varepsilon_\text{B1g}>0$ and $<0$.  

In Fig.~\ref{figResistivityAnisotropy}(a) we show the nematic resistive anisotropy at $T<T_\text{s}$, normalized by $\rho_\text{A1g}$ (with, again, an estimate for the twin boundary resistivity
subtracted; see Appendix section 6). Separate derivations from strain-ramp data from Sample C, and from data from Sample B, are shown in Appendix section 8; the agreement is excellent, which confirms
that the twin boundary resistance has been properly cancelled. The nematic resistive anisotropy peaks at $\approx$7\%, at $T \approx 80$~K, but then decreases as $T$ is reduced further, eventually
changing sign at $\approx$40~K. The low-temperature resistive anisotropy, where the nematicity is fully developed, is about $-1.5$\%. This is surprisingly small: in ARPES data, the length-to-width
ratios of the Fermi surfaces at $X$ and $\Gamma$ is 2--3~\cite{Rhodes18PRB}. Any anisotropy in conduction from these Fermi surfaces individually appears to cancel almost perfectly. In contrast,
resistive anisotropy in materials with magnetic order is much larger, for example on the order of 100\% in underdoped Ba(Fe,Co)$_2$As$_2$~\cite{Chu10Science}.

In Ref.~\cite{Tanatar16PRL}, $\rho_a$ and $\rho_b$ were obtained by comparing the resistivities of stress-detwinned and unstressed samples, taking the resistivity of the latter to be $(\rho_a +
\rho_b)/2$. $(\rho_a - \rho_b)/(\rho_a + \rho_b)$ was found to be $\approx$3\%, with weak temperature dependence, in qualitative disagreement with the results here. However, this analysis method
treats the twin boundary resistance as negligible, which we have shown not to be a good approximation at lower temperatures.

This is, however, a valid approach near $T_\text{s}$, where twin boundary resistance is low compared with the total sample resistance. We show in the inset of Fig.~\ref{figResistivityAnisotropy}(a)
$\rho_a$ and $\rho_b$ of Sample C, derived by taking the $T$-ramp resistivity at $\varepsilon_\text{B1g}=0$ as $(\rho_a + \rho_b)/2$ and then applying the anisotropy plotted in the main panel to
obtain $\rho_a$ and $\rho_b$. Upon cooling into the nematic phase, $\rho_b$ is seen to decrease and $\rho_a$ to increases.

In Fig.~\ref{figResistivityAnisotropy}(b) we compare the B\textsubscript{1g} resistivity derived from the long strain ramps shown in Fig.~\ref{figMontgomery}(a) to that from a classic
elastoresistivity measurement, also performed on Sample C, in which the strain was oscillated by a small amplitude (here, a peak-to-peak amplitude of $3.4 \times 10^{-5}$ at 0.0167~Hz) and the
resulting oscillation amplitude of the resistivity was measured. For the long strain ramps, the nematic resistive anisotropy at $T<T_\text{s}$ was determined by methods similar to those described
above (See Appendix section 8 for details), and the B\textsubscript{1g} elastoresistivity is taken as $(1/\varepsilon_\text{s}) \times (\rho_a - \rho_b)/(\rho_a + \rho_b)$. For $T>T_\text{s}$ and for
the small-amplitude strain oscillation data, the B\textsubscript{1g} elastoresistivity is $(1/\rho_\text{A1g}) \times d\rho_\text{B1g}/d\varepsilon_\text{B1g}$; these two definitions are equivalent at
$T = T_\text{s}$. Perhaps surprisingly, the small-amplitude elastoresistivity tracks the long-strain-ramp data to well below $T_\text{s}$, which shows that even with a very small strain oscillation
amplitude twin boundaries shift with the applied strain.

The B\textsubscript{1g} elastoresistivity of Sample C peaks at 62. Previously reported values, from conventional measurements in which samples are affixed directly to piezoelectric actuators, are
61~\cite{Tanatar16PRL}, 38~\cite{Hosoi16PNAS}, and 300~\cite{Watson15PRB}. We fit the small-amplitude data at $T>T_\text{s}$ to a Curie-Weiss form,
\[
\frac{1}{\rho_\text{A1g}}\frac{d\rho_\text{B1g}}{d\varepsilon_\text{B1g}} = \frac{a}{T - T_\text{s,0}},
\]
which yields $T_\text{s,0} = 60.7$~K. A similar fit to data from Sample B [where, because $\rho_{010}$ was not measured, we analyse the quantity $(1/\rho_{100})d\rho_{100}/d\varepsilon_{100}$] yields
$T_\text{s,0} = 54.8$~K.  (We note that we do not include a high-temperature offset term in these fits, because doing so returns negative values, implying that in the $T \rightarrow \infty$ limit
compression would cause resistivity to increase, which is not expected.) 

In Fig.~\ref{figResistivityAnisotropy}(c) we show the normalized resistive anisotropy of Sample C as a function of strain at $T \approx T_\text{s}$. This peaks at $\approx$6\%, at
$\varepsilon_\text{B1g} = -0.18 \times 10^{-2}$, then shrinks as $\varepsilon_\text{B1g}$ becomes more negative. In order to estimate the magnitude of the strain-induced nematicity at this strain, we
evaluate parameters in a Ginzburg-Landau free energy,
\begin{equation}
F = \frac{\alpha \times (T - T_\text{s,0})}{2}\psi^2 + \frac{b}{4}\psi^4 + \frac{c}{2}\varepsilon_\text{B1g}^2 - \lambda\varepsilon_\text{B1g}\psi.
\end{equation}
We take $\psi$ to be the splitting between the $xz$ and $yz$ orbitals at the $X$ point, which grows in an order-parameter-like fashion with cooling below $T_\text{s}$ and reaches 0.05~eV as $T
\rightarrow 0$~\cite{Shimojima14PRB, Nakayama14PRL, Suzuki15PRB}. Numerical values for each parameter are determined from experimental data, as explained in Appendix section 9. The strain dependence
of $\psi$ can then be obtained by solving $dF/d\psi = 0$ under conditions of fixed strain. Doing so and evaluating at 90~K gives the result shown in Fig.~\ref{figResistivityAnisotropy}(b). The maximum
in the resistive anisotropy is found to occur when $\psi \approx 0.025$~eV, in other words when $\psi$ is approximately half of its $T \rightarrow 0$ value. This conclusion is robust against
reasonable variation of the Ginzburg-Landau parameters. When an unstressed sample is cooled, $\psi$ reaches half its $T \rightarrow 0$ value at $\approx$80~K~\cite{Shimojima14PRB}, and so we can
conclude that resistive anisotropy is a maximum for $\psi/\psi(T \rightarrow 0) \approx 0.5$ whether $\psi$ is induced through applied strain or by allowing the sample to cool.

\section{Discussion}

We first summarize our findings. 

(1) The resistive anisotropy $(\rho_a - \rho_b)/(\rho_a + \rho_b)$ evolves nonmonotonically as nematicity $\psi$ grows, peaking at $\approx$7\% and then decreasing
[Fig.~\ref{figResistivityAnisotropy}(a)]. Both when $\psi$ grows spontaneously with cooling and when it is induced through strain at $T \approx T_\text{s}$, resistive anisotropy is maximum when
$|\psi|$ is about half its spontaneous $T \rightarrow 0$ value.  

(2) The nematic resistive anisotropy changes sign at $T \sim 40$~K, and at low temperature, where the nematicity is fully developed, it is only $\approx -1.5$\% [Fig.~\ref{figResistivityAnisotropy}(a)].

(3) At $T \approx T_\text{s}$, $\rho_\text{A1g} \equiv \frac{1}{2}(\rho_a + \rho_b)$ is a minimum when the sample is tetragonal [Fig.~\ref{figMontgomery}(c)].

(4) Below $\approx$60~K biaxial compression increases both $\rho_a + \rho_b$ [Fig.~\ref{figBiaxialAndTc}(a)] and $T_\text{c}$ [Fig.~\ref{figBiaxialAndTc}(b)], in opposition to the general expectation
that compression increases bandwidths and weaken correlations.

This data set places previous low-strain measurements~\cite{Watson15PRB, Hosoi16PNAS, Tanatar16PRL} in context of the response over a wider strain range, over which elastoresistivity is a nontrivial
function of nematicity $\psi$. It allows definitive determination of the spontaneous nematic resistive anisotropy. These results are described above, so we focus the remaining discussion on
possible microscopic origins.

We first consider whether the observed elastoresistivity is a property of the mean-field nematic state.  The nematic transition point at $\varepsilon_\text{B1g}=0$ and $T = T_\text{s}$ is a
critical point of the twinning transition [see Fig.~\ref{figPhaseDiagrams}(b)], and the fact that elastoresistivity is particularly large in its vicinity, but shrinks quickly upon moving away from
it in either temperature or strain, raises the possibility that strong elastoresistivity is a consequence of critical nematic fluctuations rather than a property of the mean-field nematic state.
However, two observations argue against this possibility. One is that for $T \approx T_\text{s}$, $\rho_\text{A1g}$ is a minimum near $\varepsilon_\text{B1g}=0$ [Fig.~\ref{figMontgomery}(c)],
whereas if critical fluctuations contributed strongly to resistivity one would expect it to be maximum. The other is that the elastoresistivity is much stronger for strain aligned with than transverse
to the principal axes of the nematicity (that is, $|d\rho_\text{B1g}/d\varepsilon_\text{B1g}| \gg |d\rho_\text{B2g}/d\varepsilon_\text{B2g}|$), as expected for mean-field nematic susceptibility. We
therefore interpret the resistivities observed here as those of the mean-field nematic state.

The effects of biaxial strain at low temperature, like the observation that the superconducting gap magnitude correlates with $yz$ orbital weight~\cite{Sprau17Science, Rhodes18PRB}, point to an
important role for the $yz$ orbital in electronic correlations. The $yz$ orbital is the only one with weight both on the $\Gamma$ and $X$ pockets, and so is thought to be the dominant contributor to
$(\pi,0)$ spin fluctuations~\cite{Onari17PRB}. Inelastic neutron scattering measurements have shown that the onset of nematicity correlates with stronger $(\pi,0)$ spin fluctuations;
Refs.~\cite{Wang16NatComm, Rahn15PRB} show that there is transfer of weight, at energies $\sim k_B T_\text{s}$ relevant for transport at $T \sim T_\text{s}$, from $(\pi,\pi)$ to $(\pi,0)$ and/or
$(0,\pi)$, while in Ref.~\cite{Chen19NatMat} it is shown that the transfer is to $(\pi,0)$ rather than $(0, \pi)$.  At low temperatures the maximum $yz$ weight on the $\Gamma$ pocket is only
20\%~\cite{Rhodes18PRB}. Biaxial compression, by weakening nematicity and increasing bandwidths, will increase this value, potentially strengthening the channel for $(\pi,0)$ spin fluctuations and
causing the increase in both resitivity and $T_\text{c}$.

We focus the rest of our discussion on the nonmonotonic dependence of the resistive anisotropy on both temperature and strain. We first point out that the sign change in $\rho_a - \rho_b$ occurs within
the inelastic component of the resistivity. A possible explanation for a sign change in resistive anisotropy is that the inelastic and elastic components of the resistivity contribute oppositely, but
balance at some temperature. At 40~K, however, the resistivity is about four times the residual resistivity (based on reasonable extrapolation of the resistivity to $T \rightarrow 0$), so for this
explanation to apply the elastic resistive anisotropy would need to be about four times the inelastic resistive anisotropy, or $\sim 28$\%. The resistive anisotropy would then grow to $\sim$28\% at
very low temperatures, in disagreement with observation that it reaches only 1--2\%.

The observed temperature dependence of the resistive anisotropy does not track thermodynamic measures of nematicity. The orthorhombicity of the unstressed lattice~\cite{Kothapalli16NatComm,
Frandsen19PRB, Wang16NatMat, Rahn15PRB, Boehmer13PRB}, the anisotropy of the magnetic susceptibility~\cite{He18PRB}, and the energy splitting between the $xz$ and $yz$ bands~\cite{Nakayama14PRL,
Suzuki15PRB} all increase in a monotonic, order-parameter-like fashion below $T_\text{s}$. Several factors could cause temperature-dependent changes in resistivity. For example, in
Ref.~\cite{Breitkreiz14PRB_1} it is found that shifting the relative importance of impurity versus spin fluctuation scattering can change the sign of the resistive anisotropy in iron-based
superconductors. It is therefore important that this nonmonotonicity is also observed when nematicity is induced at fixed temperature, showing that it is not a temperature effect alone but intrinsic
to the development of nematicity.

The importance of this observation rests on the relationship between resistive anisotropy and spin fluctuations. Spin fluctuations are found in theoretical work to dominate the resistivity at higher
temperatures~\cite{Chen10PRB, Breitkreiz14PRB_1, Breitkreiz14PRB_2, Schuett16PRB, FernandezMartin19PRB, Onari17PRB}, and in optical conductivity measurements the DC resistive anisotropy is indeed
found to track the scattering rate rather than the Drude weight~\cite{Chinotti18PRB}. In Ref.~\cite{Onari17PRB}, $(\pi,0)$ fluctuations relying on the $yz$ orbital weight were found to give $\rho_a >
\rho_b$, as observed, because on the hole pocket stronger scattering of quasiparticles with $yz$ weight suppresses conduction in the $x$ direction.  At lower temperatures, when spin fluctuations are
weak, the precise locations of nesting-driven hot spots on the Fermi surface may be decisive in determining the sign of resistive anisotropy~\cite{Fernandes11PRL, Blomberg13NatComm}, making 
it sensitive to details, but as temperature is raised the precise nesting conditions become less important~\cite{Breitkreiz14PRB_1}.

A further intuitive reason to expect $(\pi,0)$ spin fluctuations to play a strong role in transport is that they connect the $\Gamma$ and $X$ Fermi surface pockets, providing a channel for umklapp
scattering and momentum relaxation along the $k_x$ direction. In a clean lattice, momentum is ultimately transferred to the lattice through umklapp scattering. In systems with closed Fermi surfaces,
small-angle electron-phonon scattering can transfer momentum between the electrons and phonons, but does not relax the momentum of the combined system, and so does not contribute to dc resistivity. This
is seen in weakly correlated metals (where the electron-phonon term is readily observable) as a modification of the usual $T^5$ dependence for electron-phonon resistivity to exponentially activated,
with the activation energy corresponding to a phonon that connects Fermi surfaces~\cite{Gugan17PRSL, Hicks11PRL}. The fact that $\rho_a$ increases when nematicity onsets [see the inset of
Fig.~\ref{figResistivityAnisotropy}(a)], while $\rho_b$ decreases, is qualitatively consistent with the $(\pi,0)$ spin fluctuations providing a mechanism for preferential relaxation of transport
currents along $k_x$.

We propose a specific mechanism for the non-monotonic dependence of resistive anisotropy, consistent with data so far. $(\pi,0)$ spin fluctuations, and the associated resistive anisotropy, strengthen
as nematicity initially onsets and the Fermi velocity on the $yz$ sections of Fermi surface is reduced. These fluctuations then weaken as the nematicity grows further and suppresses the $yz$ orbital
weight on the hole pocket, cutting off this fluctuation channel. This is a proposal and a point for further investigation; the relative contributions of spin fluctuation strength and nematicity-driven
changes in Fermi surface shape to resistive anisotropy need to be determined. However, direct measurement of spin fluctuations under tunable lattice strain, through inelastic neutron scattering, would
be a very challenging experiment. It is nevertheless an important route to attempt because it could provide a direct test of a major class of theories of the nematicity of FeSe, in which it is
proposed to be driven by the increase in phase space that it allows for spin fluctuations~\cite{Yamakawa16PRX, Chubukov15PRB, Chubukov16PRX}. The potential challenge to these theories, if the
nonmonotonic resistive anisotropy observed here indeed correlates with nonmonotonic spin fluctuation strength, is to explain why the nematicity grows well past the point where it maximises spin fluctuation
strength.

Regardless of how that path of inquiry develops, we anticipate that the strain-tuning capabilities demonstrated here will allow resolution of the separate orbital contributions to the electronic
properties of FeSe, and theories of the nematicity of FeSe to be tested.

\section{Acknowledgements}

We thank Hiroshi Kontani, Andreas Kreisel, Kazuhiko Kuroki, Seiichiro Onari, Sahana R\"{o}\ss{}ler, J\"{o}rg Schmalian, Roser Valent\'{i}, Matthew Watson, and Steffen Wirth for useful
discussions. S.H.  and T.S. thank S. Kasahara, Y. Matsuda, K. Matsuura, and Y. Mizukami for early-stage collaboration on sample growth. We thank the Max Planck Society for financial support. C.W.H.,
A.P.M., and C.T. acknowledge support by the DFG (DE) through the Collaborative Research Centre SFB 1143 (projects C09 and A04). C.T. acknowledges support by the DFG (DE) through the Cluster of
Excellence on Complexity and Topology in Quantum Matter ct.qmat (EXC 2147).  Work in Japan was supported by Grants-in-Aid for Scientific Research (KAKENHI) (Nos. JP19H00649 and JP18H05227), and
Grant-in-Aid for Scientific Research on innovative areas ``Quantum Liquid Crystals'' (Nos. JP19H05824 and JP20H05162) from Japan Society for the Promotion of Science (JSPS).

\section{Appendix} 
\numberwithin{equation}{section}
\setcounter{equation}{0}
\renewcommand{\theequation}{A\arabic{equation}}

\subsection{1. Montgomery conversion}

To measure the resistivity  $\rho_{xx}$ parallel to the direction of applied strain in FeSe we used a four-point setup with bar-shaped samples. By applying compressive and tensile strain the resistive
anisotropy in the nematic state can be extracted. To decompose the elastoresistance into its irreducible representations, and access to the nematic susceptibility requires the
knowledge of both $\rho_{xx}$ and $\rho_{yy}$ under applied strain. 

This can be achieved, for example, by measuring two samples in perpendicular orientations, special arrangement of contact geometry with respect to sample orientation and strain direction, or by a
Montgomery-type setup which  allows the simultaneous measurement of $\rho_{xx}$ and $\rho_{yy}$ in a single sample. These different approaches have been recently reviewed in Ref. \cite{Shapiro2016}.
In our platform-based measurements, the center region of the platform is small, the strain transmission length requires the sample dimension along the longitudinal axis of the platform to exceed $\approx 200$~$\mu$m. Therefore a Montgomery configuration is more suitable.

The Montgomery method \cite{Montgomery1971,Wasscher1961} allows us to convert a sample with anisotropic resistivities $\rho_i$ but rectangular shape into an
isotropic sample, with a single $\rho$, and different effective dimensions. With a rectangular shaped sample of dimensions $L_1$, $L_2$, and thickness $L_3$, the function $H$ determines the relation
between resistivity and measured resistance $R$. Following the derivations from dos Santos \emph{et~al.}. \cite{Santos2011}, the resistivity of an isotropic sample $\rho$ and a rectangular sample with
dimensions $L_1, L_2$, thickness $L_3$ and measured resistances $R_1$, $R_2$ can be expressed as
\begin{equation}
\rho = H_1 t_\text{eff} R_1
\end{equation} where H is only a function of the geometry of the sample, i.e.
$H_1 = H(L_1, L_2)$, $H_2 = H(L_2, L_1)$ and the effective thickness $t_\text{eff}=t_\text{eff}(L_3)$. 

Now we can compare the ratios
\begin{equation}
H_1/H_2 = R_1/R_2
\label{H_ratios}
\end{equation}
which can be used to calculate $L_1/L_2$ in several ways
Either with the definition of 
	\begin{equation}
1/H_1 = 4/\pi \sum^{\infty}_{n=0} 2/\{(2n + 1)\sinh[\pi (2n + 1) L_1/L_2)] \}
	\end{equation}
	
 from \cite{VanderPauw1961}, or
using the approximation 
\begin{equation}
\frac{L_2}{L_1} \approx \frac{1}{2}\left[\frac{1}{\pi}\ln\frac{R_2}{R_1} +  \sqrt{\left[\frac{1}{\pi}\ln\frac{R_2}{R_1} \right]^2 + 4}  \right]
\end{equation} derived by dos Santos \emph{et al.}. \cite{Santos2011}.

The infinite series converges rapidly, we therefore compute the first few terms and use a bisection algorithm to solve eq. \eqref{H_ratios}.

Furthermore we require two relations from Wasscher's transformation \cite{Wasscher1961}:
\begin{equation}
L_i = L'_i \sqrt\frac{\rho_i}{\rho}
\end{equation}
and 
\begin{equation}
\rho^3 = \rho_1\rho_2\rho_3,
\end{equation}
which connects the length of an isotropic sample $L_i$ with the corresponding dimensions and resistivity of the anisotropic sample $L'_i$ and $\rho_i$. 
%
%
With the definition of effective thickness 
$t_\text{eff}' = t_\text{eff}\ (L'_3/L_3)$ and in the limit of thin samples, i.e. $L_3 / (L_1 L_2)^{1/2} < 0.5$ the ratio $t_\text{eff}/L_3 \approx 1$, and therefore also $t_\text{eff}' \approx L'_3$.

This allows us to derive 
\begin{equation}
\left(\rho_1\rho_2\right)^{1/2} = H_1 t_\text{eff}' R_1
\end{equation}
which yields a relationship between $\rho_1$ and $\rho_2$:
\begin{equation}
\rho_1 = \frac{L_2'^2}{L_1'^2} \frac{L_1^2}{L_2^2} \rho_2.
\end{equation}
$\rho_2$ is derived from the measured resistances and sample dimensions:
\begin{equation}
\rho_2  = H_1 t_\text{eff}' R_1 \frac{L_1'}{L_2'}\frac{L_2}{L_1}.
\end{equation}

\paragraph{Applied strain contribution.}
When applying uniaxial strain to a sample the apparent elastoresistance consists of a purely geometric contribution from the change of its dimensions and the strained material exhibits a different resistivity.
We take the geometric contribution into account by calculating the strained sample dimensions in the limit of small strains. 

For strains within the plane we assume that the platform is coupled rigidly enough to the sample that its dimensions follow the applied strain from the platform:
\begin{align}
L_{1,\mathrm{strained}}' = L_{1}' \, (1 + \epsilon_{xx})\\
L_{2,\mathrm{strained}}' = L_{2}' \, (1 + \epsilon_{yy})
\end{align}
The $c$-axis of the sample is not constrained in the experiment. 
If we assume almost rigid coupling within the plane, the corresponding response of the sample along the $c$-axis can become significant. To include this effect we can define a renormalized Poisson's ratio $\nu_{\perp}^*$ for the out-of-plane component:
\begin{equation}
\nu_{\perp}^* = \beta\nu_{\perp}
\end{equation}
Here $\beta$ depends on the elastic moduli and the Poisson's ratios as follows:

\begin{equation}
\beta = \frac{E_\parallel}{E_{\perp}} \, \frac{1-\nu_{\mathrm{eff}}}{1-\nu_{\parallel}}
\end{equation}

\subsection{2. Strain transmission}

When the epoxy and sample layers are both thin and the epoxy elastic moduli are low, strain transfer to the sample can be characterized to good accuracy by a strain transmission length $\lambda$,
given by $\lambda = (ctd/G)^{1/2}$, where $c$ is the relevant elastic modulus of the sample, $t$ the sample thickness, $d$ the epoxy thickness, and $G$ the epoxy shear modulus~\cite{Hicks14RSI}. Under
the conditions that the $c$-axis strain in the sample is unconstrained while the transverse strain is fixed, $c = c_{11} - c_{13}^2/c_{33}$~\cite{Park20RSI}. Even though the Young's modulus of FeSe
becomes nearly zero for $T \approx T_\text{s}$~\cite{Boehmer15PRL}, $c$ remains substantial, at $\approx$40~GPa based on the elastic moduli reported in Refs.~\cite{Zvyagina13EPL, Millican09SSC,
Margadonna09PRB}.  Physically, this means that the lattice remains stiff against biaxial compression, even as it becomes soft against orthorhombic distortion.  To determine $d$, a focused ion beam was
used to slice through some of the samples at a few points; an example of a cross section through Sample B is shown in Fig.~\ref{figSetup}(d). $d$ was found to be 5--10~$\mu$m.  To estimate $G$ we take
the Young's modulus of Stycast 1266, reported in Ref.~\cite{Hashimoto80RSI}, and assume a Poisson's ratio of 0.3, which gives $G = 1.6$~GPa at low temperature. 

Samples A and B are both long, ensuring good coupling of longitudinal strain to the platform, and so the key question is of their width in comparison with $\lambda$. For samples much narrower than
$\lambda$, the transverse strain is the longitudinal strain multiplied by the sample's Poisson's ratio, while for samples much wider than $\lambda$, it is the longitudinal strain multiplied by the
platform's Poisson's ratio. For FeSe this is an important distinction because its Poisson's ratio for $T \sim T_\text{s}$ is close to 1, while that of titanium is 0.32.  We find that all of the
samples have a width larger than $\approx 4\lambda$, ensuring good locking of both longitudinal and transverse strains to the platform. In particular, Sample A is 31~$\mu$m thick, yielding $\lambda
\approx 60$~$\mu$m, while its width is 280~$\mu$m. Sample B is 10~$\mu$m thick, yielding $\lambda \approx 40$~$\mu$m, and 230~$\mu$m wide. Complete sample dimensions are shown in
Table~\ref{tab:addlabel}.

\begin{table}[htbp]
        \centering
	\caption{Sample parameters: length, width, thickness, separation $l_\text{contact}$ of the voltage contacts, and the residual resistivity ratio $\rho(\text{300 K})/\rho(\text{12 K})$. Note
that at 12~K there is still strong inelastic scattering.}
        \begin{tabular}{llllll}
                \toprule  
		Sample \hspace{1mm} 	& $l$~($\mu$m)\hspace{1mm} 	& $w$~($\mu$m)\hspace{1mm} 	& $t$~($\mu$m)\hspace{1mm} 	& $l_\text{contact}$ ($\mu$m)\hspace{1mm}	&  RRR \\
		\colrule
		A			& 2370				& 280				& 31				& 970						& 26 \\
		B			& 1150				& 230				& 10				& 630						& 22 \\
		C			& 434				& 425				& $\approx$10			&						&    \\
                \botrule  
        \end{tabular}%
        \label{tab:addlabel}%
\end{table}

\subsection{3. Elastic moduli}

Ref.~\cite{Zvyagina13EPL} gives elastic moduli of FeSe at $T \approx T_\text{s}$: $c_{11} \approx c_{12} \approx 50$~GPa, $c_{33} \approx 40$~GPa, and $c_{13} \approx 20$~GPa. Under conditions of
hydrostatic pressure, $\sigma / \varepsilon_{xx} = (c_{11} c_{33} + c_{12} c_{33} - 2c_{13}^2)/(c_{33} - c_{13})$, where $\sigma$ is the applied stress, and $\varepsilon_{zz} / \varepsilon_{xx} =
(c_{11} + c_{12} - 2c_{13})/(c_{33}
- c_{13})$. Under conditions of in-plane biaxial stress, where $\sigma_{xx} = \sigma_{yy}$ and $\sigma_{zz}=0$, $\sigma_{xx} / \varepsilon_{xx} = c_{11} + c_{12} - 2c_{13}^2/c_{33}$, and
  $\varepsilon_{zz}/\varepsilon_{xx} = -2c_{13}/c_{33}$.

\subsection{4. Plastic deformation of the platform}

\begin{figure}[tbhp]
	\includegraphics[width=0.95\linewidth]{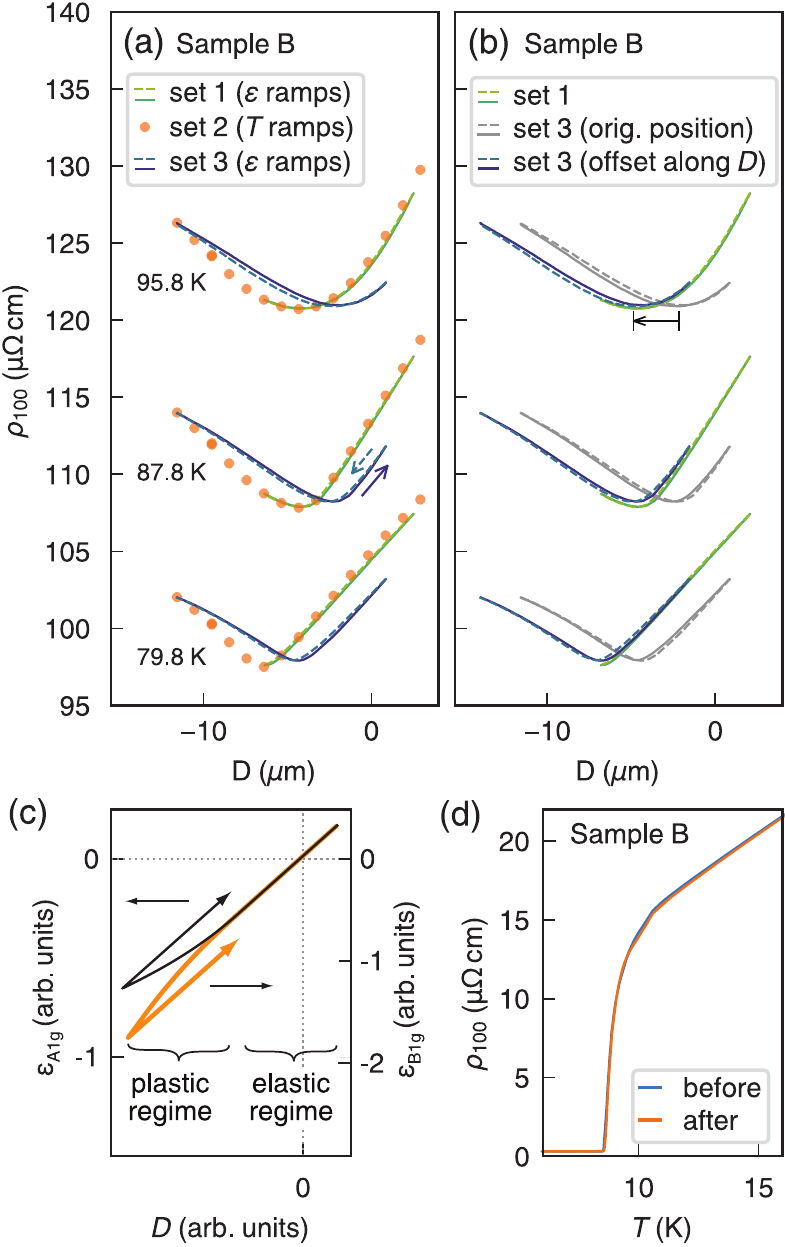}
	\caption{\figPlasticCaption{}}
	\label{figPlastic}
\end{figure}

Sample B was driven to high compressions, and the platform deformed plastically when the displacement $D$ applied to it exceeded $\approx7~\mathrm{\mu m}$, causing the strain in the neck to exceed the
elastic limit of the platform material, $\approx 2 \times 10^{-3}$. Data from Sample B were taken in the following order: (1) Strain ramps were performed at $T \approx T_\text{s}$ up to modest
strains. (2) Temperature ramps were performed at constant strain, incrementing the strain at 103.7~K, and moving gradually to high compressions.  (3) Further strain ramps were performed at high
compression.  Data from these three sets are plotted against $D$ in Fig.~\ref{figPlastic}(a--b). There is low hysteresis within each strain ramp data set, and the two strain ramp data sets match
closely except for an offset along the $D$ axis. The temperature ramp data bridge this offset smoothly. We conclude that the platform deformation was essentially elastic within each strain ramp data
set, and that the offset between them is due to plastic deformation caused by the large change in applied strain over the course of the temperature ramps.

Fig.~\ref{figPlastic}(c) shows a schematic illustration of the expected form of the plastic deformation. Initially, when the platform deformation is elastic, $\varepsilon_{\text{A1g}}$ and
$\varepsilon_{\text{B1g}}$ are linear in $D$: $\varepsilon_{\text{B1g}} = 0.66D/l_\text{eff}$ and $\varepsilon_{\text{A1g}} = 0.34D/l_\text{eff}$ (where $l_\text{eff}$ is the effective length of the
platform). Beyond its elastic limit, the platform material resists further volume compression by flowing plastically outward: $\varepsilon_{\text{B1g}}$ starts to vary more steeply with $D$, and
$\varepsilon_{\text{A1g}}$ less steeply. When the direction of the applied displacement is reversed, the platform deformation is again elastic over some range, but for a given $D$
$\varepsilon_{\text{B1g}}$ is larger and $\varepsilon_{\text{A1g}}$ smaller than before. 

That the sample deformation remained elastic even as the platform deformed plastically is shown in Fig.~\ref{figPlastic}(d), in which low-temperature data from before and after the plastic
deformation, taken at strains where $T_\text{c}$ is the same, are plotted together. The residual resistivity is unchanged.

The sign of the offset between the pre- and post-plastic-deformation data shows that $\rho_{100}$ is controlled dominantly by $\varepsilon_{\text{B1g}}$, rather than $\varepsilon_\text{A1g}$. The fact
that a horizontal displacement works so well to match the pre- and post-plastic deformation data shows that the effect of $\varepsilon_{\text{A1g}}$ on $\rho_{100}$ is small; if it were strong then it
would have to be finely balanced, over a wide temperature range, with that of $\varepsilon_{\text{B1g}}$ for the net effect to be so neatly a horizontal shift of the $\rho(D)$ curves.  Furthermore,
the data of Fig.~\ref{figSampleA} show directly that the dependence of $\rho$ on $\varepsilon_{\text{A1g}}$ is weak.

In Fig.~\ref{figSampleBComplete}(a) and (c), to account for this plastic platform deformation data from the high-strain strain ramps are offset by $\Delta \varepsilon_\text{B1g} = -0.072 \times
10^{-2}$. Because this deformation occurred gradually over the course of the temperature ramps, for $\varepsilon_{\text{B1g}} < -0.11 \times 10^{-2}$ each individual temperature ramp is offset along
the $\varepsilon_{\text{B1g}}$ axis to match the resistivity at 103.7~K with that from the strain ramps.

\subsection{5. Annealing twin boundaries}

\begin{figure}[htpb]
\includegraphics[width=0.99\linewidth]{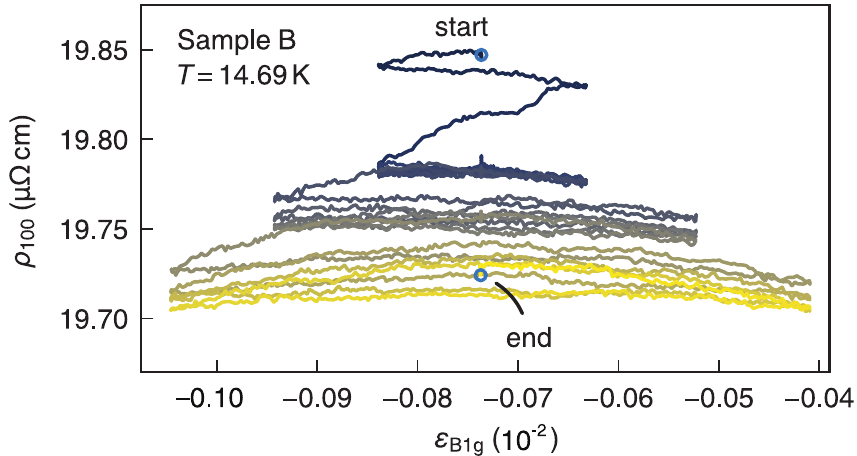}
\caption{\figTwinAnnealingCaption{}}
\label{figTwinAnnealing}
\end{figure}

In Fig.~\ref{figTwinAnnealing} we show results of a twin boundary annealing experiment. Sample B was cooled from above $T_\text{s}$ to 14.69~K at a fixed strain. The strain was then ramped back and
forth. Over the first few cycles of strain ramping, the sample resistance falls, but then settles at a lower value. When the strain ramp amplitude is then increased, the decrease in resistance
resumes, and then the resistance settles at a yet lower value. This behavior shows that twin boundaries can be partially annealed out of the sample through strain ramps, and confirms that the peaked
form of the resistance in $T$-ramp data, shown in Fig.~\ref{figSampleBComplete}, is due to twin boundaries.

\subsection{6. \boldmath{$T$}-ramp data from Sample C, and twin boundary resistivity}

\begin{figure}[htbp]
	\includegraphics[width=0.99\linewidth]{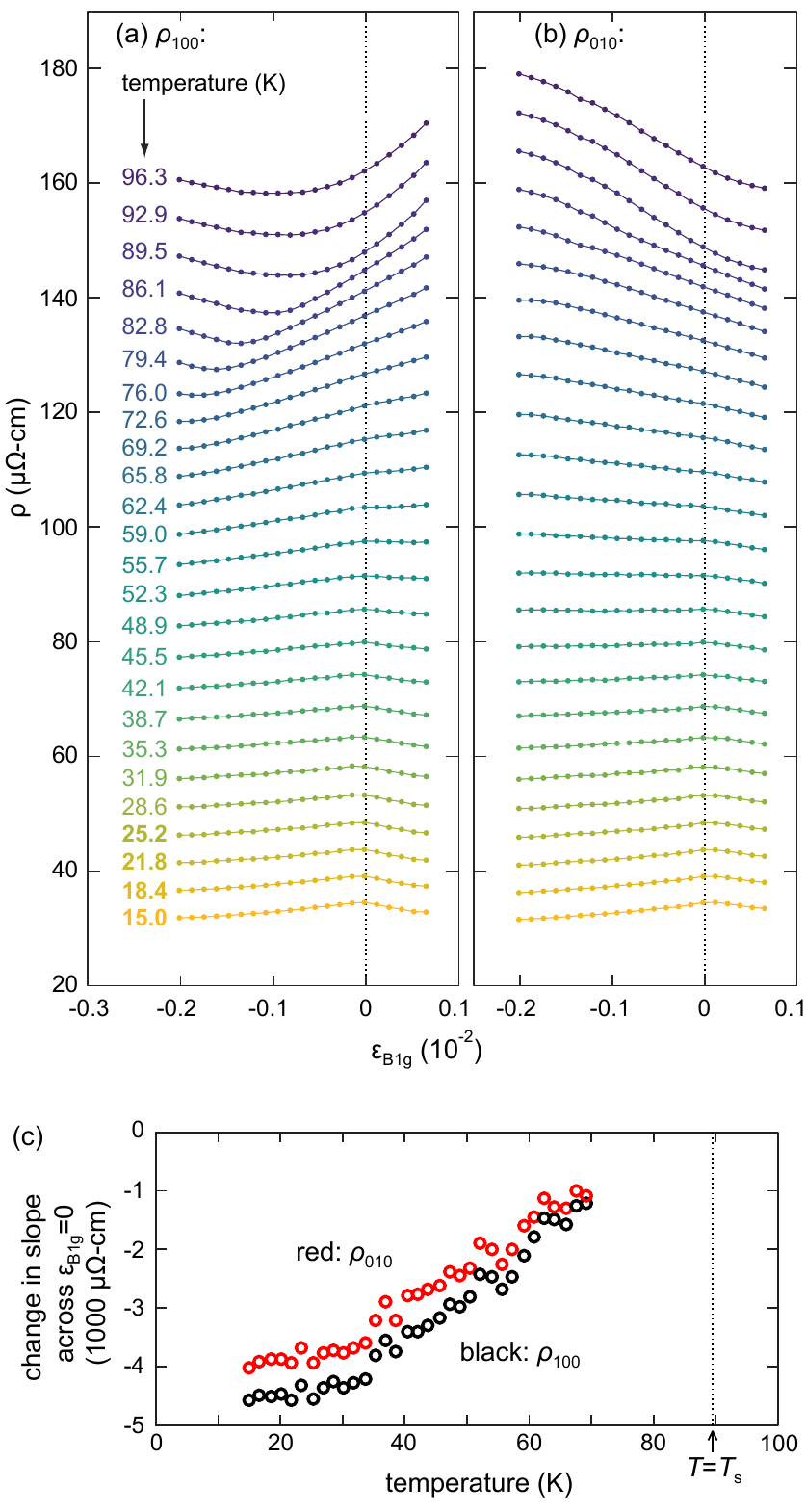}
	\caption{\figSampleCTempRamps{}}
	\label{figSampleCTempRamps}
\end{figure}

Temperature-ramp data from Sample C are shown in Fig.~\ref{figSampleCTempRamps}. At low temperatures, the cusp in $\rho(\varepsilon_\text{B1g})$ due to the maximum in domain wall density is visible in
both $\rho_{100}$ and $\rho_{010}$. Its location differs slightly in the two measurements, possibly because in the Montgomery configuration measurements of $\rho_{100}$ and $\rho_{010}$ do not probe
precisely the same area of the sample. We take $\varepsilon_\text{B1g}=0$ as the average of the cusp locations in $\rho_{100}$ and $\rho_{010}$.

In Fig.~\ref{figSampleCTempRamps}(c), we show the change in slope $d\rho/d\varepsilon_\text{B1g}$ across the cusp at $\varepsilon_\text{B1g}=0$ versus temperature. This quantity is proportional to the
twin boundary contribution to sample resistivity at $\varepsilon_\text{B1g}=0$. The twin boundary resistivity is seen to be nearly $T$-independent up to $\sim$30~K, and then to decrease. Note that
this is the twin boundary resistivity when the sample is cooled from above $T_\text{s}$ at $\varepsilon_\text{B1g}=0$; when it is brought to $\varepsilon_\text{B1g}=0$ by ramping strain at constant
temperature, the twin boundary density is lower.

In Figs.~\ref{figBiaxialAndTc}(a) and~\ref{figResistivityAnisotropy}(a), elastoresistivities normalized by $\rho_a + \rho_b$ are shown. For this normalization we subtracted off an estimated twin
boundary resistivity, $\rho_\text{TB}(T)$; for example, in Fig.~\ref{figResistivityAnisotropy}(a) the quantity that is plotted is $\rho_a - \rho_b$, determined by the underlying slopes method described in the text, divided by
$\rho_{100}(\varepsilon_\text{B1g}=0) + \rho_{010}(\varepsilon_\text{B1g}=0) - 2\rho_\text{TB}(T)$. Based on the illustration in Fig.~\ref{figSampleBComplete}(c), we estimate $\rho_\text{TB}(T
\rightarrow 0) = 3$~$\mu\Omega$-cm. We take $\rho_\text{TB} = \rho_\text{TB}(T \rightarrow 0) \times [1 - (T/T_\text{s})^2]$. This form overestimates somewhat the true twin boundary resistance as $T$
approaces $T_\text{s}$, however the effect is tiny.

\subsection{7. Elastoresistivity of Sample A}

\begin{figure}[htbp]
	\includegraphics[width=0.99\linewidth]{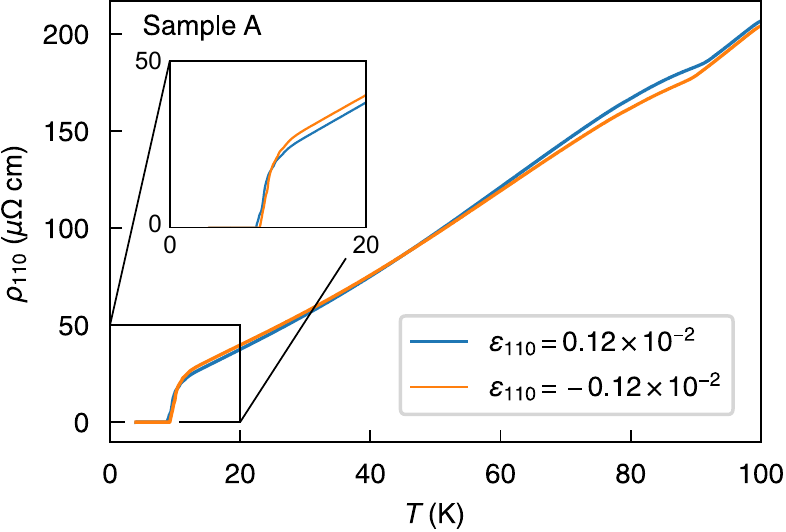}
	\caption{\figAppendixBCaption{}}
	\label{figAppendixB}
\end{figure}

Fig.~\ref{figAppendixB} shows the elastoresistivity of Sample A over a wide temperature range. The behavior qualitatively matches the A\textsubscript{1g} elastoresistivity determined from Sample C,
and plotted in Fig.~\ref{figBiaxialAndTc}(a): at higher temperatures, compression causes a decrease in resistivity, and at lower temperatures an increase. The sign of the response changes at $T
\approx 45$~K, against 60~K for the A\textsubscript{1g} elastoresistivity of Sample C. The measured resistivity of Sample A will also be affected by the B\textsubscript{2g} elastoresistivity; however,
because this is transverse to the nematic axes it is not expected to be large, and the qualitative agreement with the A\textsubscript{1g} elastoresistivity suggests that it is indeed much smaller than
the A\textsubscript{1g} elastoresistivity. Note also that $T_\text{c}$ increases with compression, as observed in Sample B [Fig.~\ref{figBiaxialAndTc}(b)].

\subsection{8. Additional derivations of the nematic resistive anisotropy}

\begin{figure}[htbp]
	\includegraphics[width=0.99\linewidth]{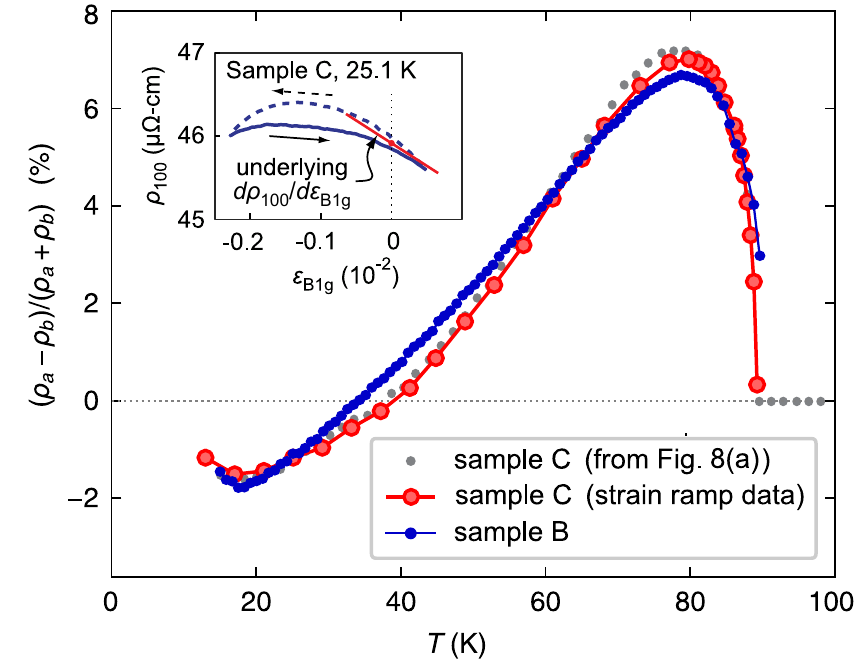}
	\caption{\figAppendixACaption{}}
	\label{figAppendixA}
\end{figure}

Above, we presented a determination of the nematic resistive anisotropy for $T<T_\text{s}$ based on Sample C temperature-ramp data, in which the twin distribution can be assumed to be in near
equilibrium with the applied strain. Here, we analyze strain-ramp data. As described above, the determination of nematic resistive anisotropy depends on extraction of the slopes
$d\rho_{100}/d\varepsilon_\text{B1g}$ and $d\rho_{010}/d\varepsilon_\text{B1g}$ at $\varepsilon_\text{B1g} = 0$ and under the condition that the twin boundary configuration does not change. In the
strain ramps, the density and location of twin boundaries lags the applied strain, and we therefore obtain these slopes by averaging the observed slopes from the increasing-strain and
decreasing-strain ramps, as illustrated in the inset of Fig.~\ref{figAppendixA}. Applying Eq.~(6) yields the nematic resistive anisotropy plotted in Fig.~\ref{figAppendixA}. The close agreement with
$T$-ramp data shows that the twin boundary resistance has been properly excluded. Note that, because the twin boundary density is lower in strain-ramp than temperature-ramp data, we do not subtract
off a twin boundary contribution.

Also shown in Fig.~\ref{figAppendixA} is the resistivity anisotropy determined from Sample B. For Sample B, only $\rho_{100}$ was measured. Evaluating Eq.~(1) at $f = 0.5$ yields
\begin{equation}
\rho_a - \rho_b = 2\varepsilon_\text{s} \left( \frac{d\rho_{100}}{d\varepsilon_\text{B1g}} -
\frac{1-\nu}{1+\nu}\frac{d\rho_\text{A1g}}{d\varepsilon_\text{A1g}}\right).
\end{equation}
$d\rho_\text{A1g}/d\varepsilon_\text{A1g}$ must be taken from data from Sample C [see Fig.~\ref{figBiaxialAndTc}(a)]; the data plotted in Fig.~\ref{figAppendixA} includes this correction. For the
normalization our estimate for twin boundary resistivity is subtracted (see Appendix section 6).

\subsection{9. Ginzburg-Landau parameters}

In the Ginzburg-Landau free energy [Eq.~(7)], the strain is the B\textsubscript{1g} strain, for which the elastic constant $c$ is $c_{11} - c_{12}$. This elastic constant must be evaluated without the
influence of nematic susceptibility.  Ref.~\cite{Zvyagina13EPL} finds $c_{11} \approx 80$~GPa at $T \approx 250$~K, and electronic structure calculations give $c_{11} =
95$~GPa~\cite{Chandra10PhysicaC}. We take the estimate $c = c_{11} - c_{12} = 60$~GPa.  The structural strain is obtained by noting that $dF/d\varepsilon = 0$ at $\varepsilon = \varepsilon_\text{s}$,
which gives $\varepsilon_\text{s} = (\lambda/c)\psi$. Although the Ginzburg-Landau formalism only applies, strictly, very near to $T_\text{s}$, we evaluate parameters at considerably lower temperature
in order to obtain approximate evaluations of the coefficients.  $\varepsilon_\text{s} \rightarrow 0.27 \times 10^{-2}$ as $T \rightarrow 0$~\cite{Kothapalli16NatComm}, yielding a value for the
coupling constant: $\lambda \approx 3.2$~GPa/eV. As shown in fig.~\ref{figResistivityAnisotropy}(b), a fit to elastoresistivity data yields a bare nematic transition temperature $T_\text{s,0} =
60.7$~K; we take $T_\text{s,0} = 60$~K. $T_\text{s}$ is defined by the relationship 
\begin{equation} 
\frac{\lambda^2}{c} - \alpha \times (T_\text{s} - T_\text{s,0}) = 0.  
\end{equation} 
Taking $T_\text{s} = 90$~K yields $\alpha = 0.0057$~GPa/eV$^2$-K. Finally, we evaluate $b$ from the observation that $\psi$ reaches half its $T \rightarrow 0$, or 0.025~eV, value at $T \approx
0.9T_\text{s}$~\cite{Shimojima14PRB}, which gives $b = 82$~GPa/eV$^4$.

\bibliography{literature_FeSe.bib}

\end{document}